# Weighing votes in human-machine collaboration for hazard recognition: Inferring hazard perceptual threshold and decision confidence from electroencephalogram wavelets

Xiaoshan Zhou[(1)], Pin-Chao Liao[(1)]*

[(1)]Department of Construction Management, Tsinghua University, No. 30, Shuangqing Rd., HaiDian District, Beijing 100084, PR China.

*Corresponding author: pinchao@tsinghua.edu.cn

## Abstract

Purpose

The implementation of human-machine collaboration is a promising strategy to improve on-site hazard recognition for hazard inspection. However, research on the effective integration of opinions from humans with machines for optimal group decision making is lacking. Hence, considering the benefits of a brain-computer interface (BCI) to enable intuitive commutation, this study proposes a novel method to predict human hazard response choices and decision confidence from brain activities for a superior confidence-weighted voting strategy.

Design/methodology/approach

First, we developed a Bayesian inference-based algorithm to ascertain the decision threshold above which a hazard is reported from human brain signals. This method was tested empirically with

**Abbreviations**:
BCI: brain–computer interface; SDT: signal detection theory; ROIs: regions of interest; LPA: latent profile analysis; BIC: Bayesian information criterion; MAP: maximum-a-posteriori; ROC: receiver operating characteristic electroencephalogram (EEG) data collected in a laboratory setting and cross-validated using behavioral indices of the signal detection theory. Subsequently, based on numerical simulations, the decision criteria for low-, medium-, and high-confidence level differentiations characterized by parietal α-band EEG power were determined.

Findings

The investigated hazard recognition task was described as a process of probabilistic inference involving a decision uncertainty evaluation. The results demonstrated the feasibility of EEG measurements in observing human internal representations of hazard discrimination. Moreover, the optimal criteria to differentiate between low-, medium-, and high-confidence levels were obtained by benchmarking against an optimal Bayesian observer.

Originality/value

This study developed a novel methodical framework for decoding perceptual thresholds for reporting hazards, thus enabling the prediction of hazard responses in hazardous environments from human participants' EEG-measured brain activities. This is also the first study to explore methods to estimate decision confidence and uncertainty from neural activities in the construction context. This research demonstrates the potential of a BCI as an effective channel for intention prediction and telecommunication, laying the foundation for the design of future hazard detection techniques in the collaborative human-machine systems research field.

## 1 Introduction

The high accident rate in the construction industry is a concern for both workers and researchers (Statistics, 2019). In particular, failure of workers to identify hazards has been recognized as a high-priority precursor of accidents (Pereira et al., 2018); therefore, improving hazard recognition is the first step toward proactive accident prevention (Pandit et al., 2019). Although previous studies have addressed the inadequate hazard recognition abilities of workers, more than 50% of hazards remain unidentified (Albert et al., 2017; Sun & Liao, 2019). This is mainly due to the inherent limitations of human cognitive functions (Hodgetts et al., 2017), such as inattentional blindness (Chen et al., 2016), fatigue (Li et al., 2020), and stress (Jebelli et al., 2018).

To address the limitations of the human visual system, human–machine collaboration have been actively explored for hazard inspection (Liang et al., 2021). This concept stems from the emergence of various advanced computer vision technologies for continuous hazard monitoring, diagnoses, and warning support on construction sites (W. Liu, Meng, et al., 2021; Mostafa & Hegazy, 2021; Paneru & Jeelani, 2021). Although these technologies are becoming increasingly autonomous, their robustness is largely influenced by photometric environmental conditions (e.g., poor lighting, long distances, high speeds, and similarities with the background of objects) (Fang et al., 2020; Koch et al., 2015). Therefore, humans continue to play a central role in effective hazard identification and response (Schaefer et al., 2016).

Although previous studies proposed multiple forms of human-machine collaboration for hazard inspection, a critical problem remains unsolved. That is, when working with machines, the problem of effectively integrating opinions from both intelligent agents (human participants and automated algorithms) to form an optimal final decision. As suggested by previous studies, strategies for opinion integration can significantly influence the decision-making performance of a group (Baharad et al., 2011). The simplest and most commonly-used strategy in scenarios involving binary determinations (such as the hazard recognition task in this study, making a judgment between "hazardous' and "safe") is the simple "majority rule" strategy, which has been proven suboptimal in multiple cases (Baharad et al., 2011). Therefore, a method with minimized communication efforts and maximized group decision-making performance is advantageous for the successful implementation of human-machine collaboration in hazard inspection.

A brain–computer interface (BCI) provides a direct communication channel between the human brain and the outside world. It has been widely investigated for mental state monitoring for construction hazard prevention (Hwang et al., 2018; Tehrani et al., 2021), and a wide application of wearable BCIs is expected to be endorsed at construction sites for construction workers' cognitive status monitoring, performance analysis, and hazard warning (Abuwarda et al., 2022; Saedi et al., 2022). BCIs have also been explored for intention prediction during various cognitive tasks (Varbu et al., 2022). Therefore, we propose a wearable BCI-based method for decision integration to replace the traditional verbal communication or button-press methods. In addition, previous studies have highlighted the "intuitive" communication enabled by BCI, which implies that individuals' intentions can be directly decoded from their brain activities and converted into output commands to external devices (Y. Liu, Habibnezhad, et al., 2021). This provides the most energy-efficient means for communication independent of additional behaviors and communication efforts. Furthermore, this method enables individuals to focus on their work, which has safety advantages. Moreover, psychological measures are relatively objective because when individuals self-report decisions, they are easily affected by emotions and pressure from their peers (Han et al., 2019).

Despite the benefits of a BCI system, there is no direct evidence that the hazard perception and decision-making uncertainty of construction workers can be reliably predicted and estimated from their brain activities. To address this gap, this study proposes a novel approach for determining human decision criteria for reporting hazards and estimating decision confidence using brain wavelets during hazard recognition. Hazard recognition is a high-level cognitive process with sensory and decision-related cortical areas jointly collecting sensory cues, selectively addressing potential hazardous cues, confirming these cues by matching them with cognitive hazard representations, and responding appropriately (Kowalski-Trakofler & Barrett, 2003). In this study, we designed our approach based on signal detection theory (SDT) under Bayesian inference and previously identified neuropsychological evidence regarding hazard perception and decision-making.

*Hypothetical brain activities using SDT for a hazard recognition task*

SDT is an extensively used theory developed under the Bayesian inference framework (Pouget et al., 2013) for binary classification tasks. It provides a rigid framework to classify the attributions of individual differences in hazard recognition performance into sensitivity (ability of workers to differentiate between hazardous and safe scenarios) and response bias (threshold of perceived level of hazard above which workers respond) (Swets, 2001). SDT defines the process of hazard recognition as the process whereby the human brain continues to extract relevant information and collect evidence in favor of one of the two options (hazardous or safe) until a decision is triggered upon reaching the threshold (Ratcliff & McKoon, 2008). When a signal (the "hazard") is present, the integrated evidence is greater than when the signal is absent. As SDT curves represent hypothetical brain activity (Wickens, 2000), hazardous stimuli, opposed to safe stimuli, are expected to produce more neural activity signals. As expected, previous studies have identified the corresponding neuropsychological correlates that cause an increase in brain activity in the frontal and parietal areas during hazard processing compared with processing in safe environments (Q. Liu, 2018; Q. Ma et al., 2014; Qin & Han, 2009). Furthermore, because of the conceptual similarity between the sensitivity of SDT, (which measures the distance between the mean value of the signal and noise curves) and the brain power difference (which computes the difference in brain responses induced by hazardous and safe stimuli) we hypothesized that high-sensitivity participants demonstrated significant differences in brain activity induced by hazardous stimuli, compared to that induced by safe stimuli, whereas non-significant power differences were observed in low-sensitivity participants.

*Predictive neural activities for decision-making*

Since the sensory information perceived by individuals is typically partially informative and insufficient to uniquely comprehend the environment, inferences for hazards are probabilistic; that is, they are associated with a degree of uncertainty (Zizlsperger et al., 2016). This decison uncertainty and confidence substantially modulates neural activity (Desender et al., 2021; Gold & Shadlen, 2007; Kable & Glimcher, 2009). For example, Kiani and Shadlen (2009) demonstrated that parietal cortex activity in monkeys is associated with decision formation with an underlying degree of certainty. In addition, cortical recordings obtained through EEG measurements, such as the P300 event-related potential and error-related negativity (Navarro Cebrian et al., 2013; Selimbeyoglu et al., 2011), have been utilized to provide brain activity pattern information for decision confidence estimation. Boldt and Yeung (2015) present well-characterized neural correlates of error awareness, which indicate decision confidence, and suggested that subjective certainty is closely correlated with objective success. Desender et al. (2021) found that the magnitude of a positive-approaching centroparietal potential is generally large when the preceding choice is low confidence, which reflects the accumulation of evidence for an error. Based on these neurophysiological findings, Salazar-Gomez et al. (2017) developed a closed-loop human–robot interaction (toward a real-time intuitive interaction) using EEG-measured error-related potentials.

Regarding the correlations between brain rhythms and decision confidence, Donner et al. (2009) found that β-frequency range (12–30 Hz) activities in the motor cortex can predict the perceptual choices of participants in a "yes or no" visual detection task, before their overt manual responses, and also suggested that the motor plans for both "yes" and "no" choices result from continuously accumulating sensory evidence. O'Connell et al. (2012) demonstrated that the β-band (16–30 Hz) activity of the contralateral premotor cortex tracks cumulative evidence during decision formulation and exhibits predictive decision dynamics before a manual button press response. Kubanek et al. (2015) found that α-activity in the parietal cortex strongly encoded the confidence level of a subject for a forthcoming button-press choice. Previous research has also shown that the activity of the parietal cortex mediates between perception and action in decision-making tasks (Churchland et al., 2008; Platt & Glimcher, 1999; Shadlen & Newsome, 2001) and is closely correlated with response confidence (Kiani et al., 2014; Kiani & Shadlen, 2009). Therefore, as suggested by the aforementioned studies, the parietal cortex is a critical brain area specialized in spatially selective attention for integrating sensory information (Shafritz et al., 2002) in perceptual decision-making tasks. In this study, we chose the oscillatory activity from the parietal cortex as a proxy for decision confidence to achieve a graded evaluation of the degree of confidence characterized by EEG wavelets for a hazard recognition task.

With the attempt to infer construction worker's hazard perception threshold and decision confidence from their brain activities during hazard recognition, the main contributions of this study are as follows:

(i) A conceptual model was developed for hazard recognition under the Bayesian inference framework, and underlying hypothetical brain activities in the model were empirically examined using EEG data collected in laboratory settings.

(ii) Cross-validation was performed using the behavioral indices of SDT to prove the feasibility of electroencephalogram (EEG) measurements to observe the internal representations of humans for binary discrimination between hazardous and safe construction scenarios.

(iii) The optimal biomarkers for hazard perception were identified, benchmarking was conducted against optimal decision-making, and graded evaluation criteria for decision confidence characterized by the EEG power prior to responses were determined. These findings can be used to develop practical BCI systems to assist in decision-making processes.

## 2 Methodology

### 2.1 Participant

A total of 77 construction workers were recruited for this study. One participant was excluded because of excessive artifacts in EEG signals. Only individuals working on-site were included; therefore, individuals working as project managers were excluded. In addition, a validation test was designed for all participants (see "Stimuli and experimental protocol" for details), based on which, five participants were considered unreliable. The final sample consisted of 70 male participants (mean age = 42.2 years; age range,21–60 years). All the participants were right-handed and had normal or corrected-to-normal vision. All participants voluntarily signed informed consent forms to participate in the experiments and received 100 RMB as monetary compensation for their participation.

### 2.2 Stimuli and experimental protocol

The stimuli were provided solely by an in-use construction safety management platform (Xu et al., 2019). Hazardous construction scenes were detected during the safety inspection by construction inspectors who recorded images of hazardous conditions on the platform. After rectification, images of the corresponding corrected scenes were uploaded. The experiments included 60 construction scenes, with each scene comprising two opposing conditions (hazardous and safe). Thus, 120 trials were conducted.

An image-based hazard-recognition task was developed. In each trial, participants viewed a construction scene displayed on a computer screen and determined if it was hazardous or safe. The images were presented using Tobii Pro Lab software. To alleviate fatigue and learning curve effects, the 120 trials were randomly divided into four blocks of equal size and a 1 min break was imposed after each block.

As illustrated in Figure 1, a fixation point appeared on the screen for 500 ms at the beginning of each trial. Thereafter, a construction scene image (e.g., an elevator with an open door) is presented for up to 3000 ms, followed by a blank screen for 500 ms. The participants are subsequently required to respond to the construction scene by pressing a key on the keyboard (0 for “safe,” and 1 for “hazardous,” in Chinese). If the participants had already made a determination during the 3000 ms picture display period, they could press any key to end the image display early. The response screen was presented until the participants responded.

Finally, a validation block was implemented in which the participants responded to 30 trials randomly selected from the 120 trials. The consistency of the responses to the same stimulus (image) was verified. Five participants were excluded from the data analysis because their inconsistency rate was >50%. The entire experimental session was completed within approximately 14 min.

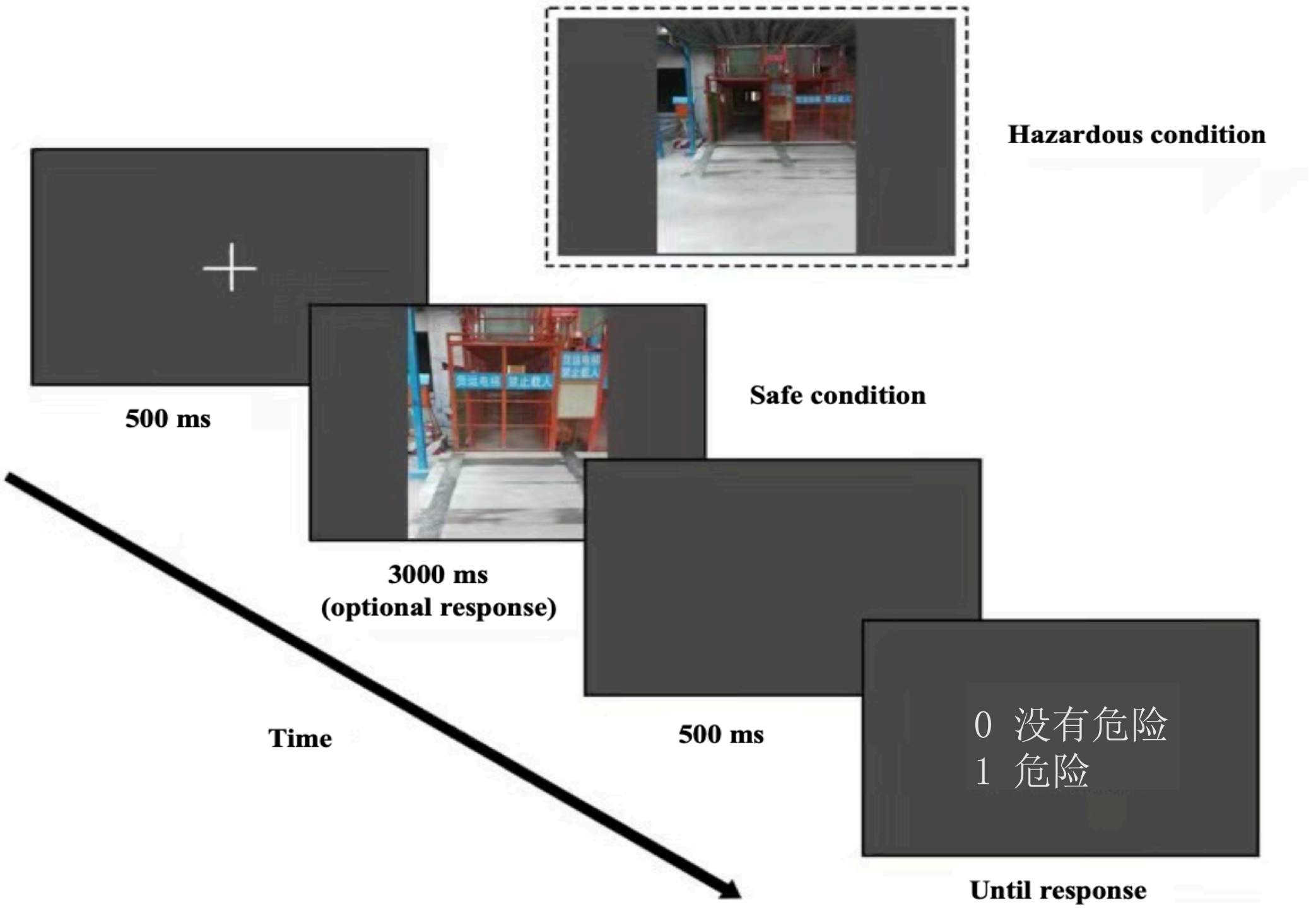

**Figure 1** Experiment paradigm (Wang & Liao, 2021).

### 2.3 EEG data recording and preprocessing

A 32-channel electrode cap (Neuroscan system, China, and Brain Products, United States) was used to record EEG signals at a sampling rate of 250 Hz. The electrodes were arranged according to the International 10–20 system (Figure 2), and the impedance of each electrode was maintained at less than 20 kΩ.

Offline analyses of EEG data were conducted using the FieldTrip toolbox (Oostenveld et al., 2011) in MATLAB (version R2019a). The EEG signals were initially subjected to band-pass filtering (0.1–40 Hz), following which possible artifacts were assessed and removed using independent component analysis (Makeig et al., 1996). The corrected data of each trial were segmented into two 1200 ms epochs, with a 200 ms pre-stimulus baseline, which began 200 ms before the stimulus onset and continued for 1000 ms after the onset. The response-locked epoch began 1000 ms prior to the response of the participants and continued for 200 ms after their response. Stimulus-locked epochs were associated with a "hazardous" or a "safe" category, depending on whether the image shown in the trial was hazardous or safe. By contrast, the response-locked epochs were associated with a "hazardous" or a "safe" category depending on whether the decision made by the participant in that trial deemed the scenario as hazardous or safe.

Furthermore, epochs with values exceeding ±100 μV at any recording electrode were rejected to avoid possible artifact contamination.

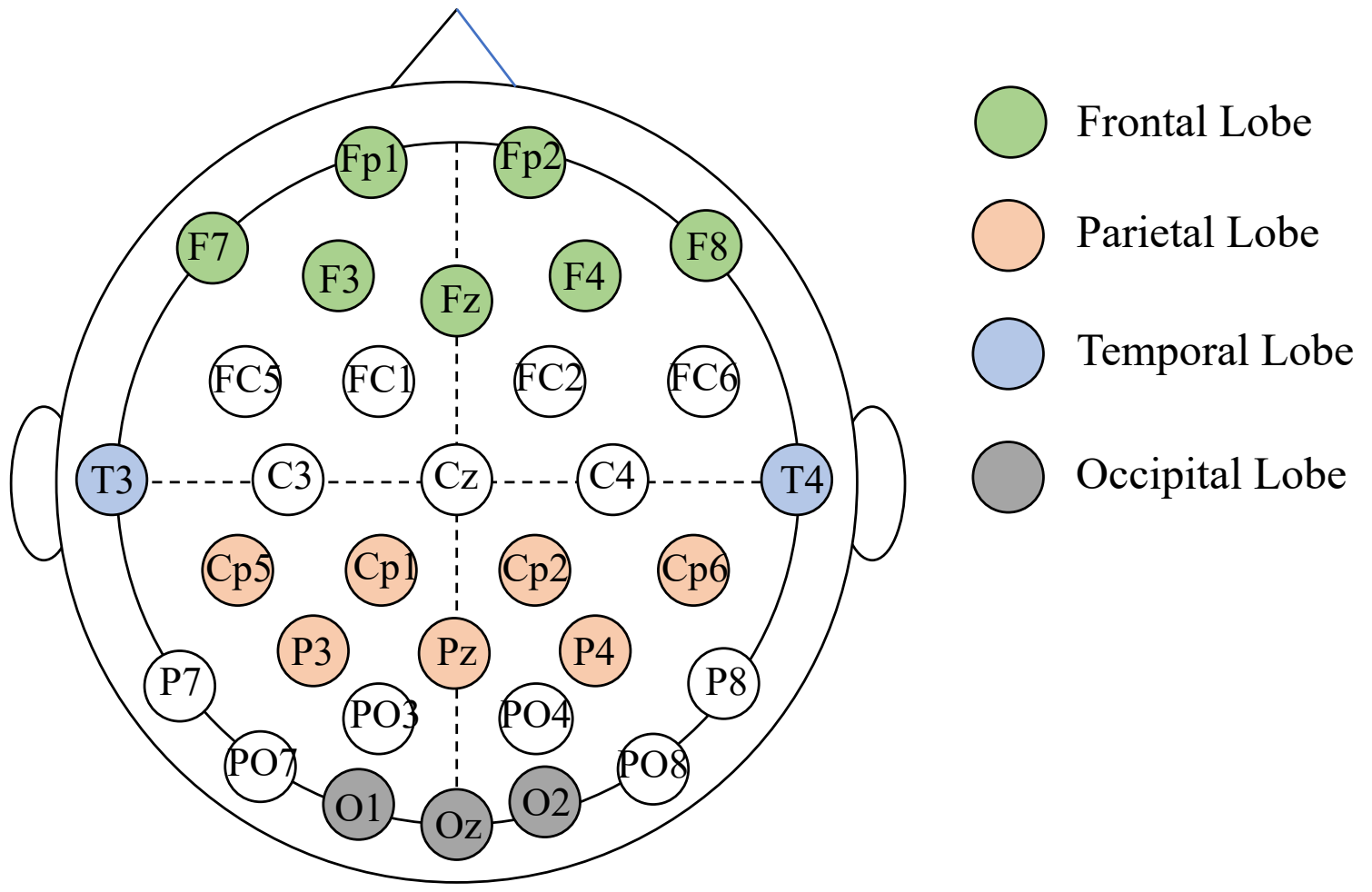

**Figure 2** Configuration of 32 electrodes

## 2.4 Time–frequency analysis of EEG data

Time–frequency analyses of the EEG data were performed using the FieldTrip toolbox (Oostenveld et al., 2011) in MATLAB. Each epoch was transformed in the frequency domain, and the time–frequency representation (denoted hereafter as "power," as suggested by Cohen and Gulbinaite (Cohen & Gulbinaite, 2013) was obtained under each condition (hazardous or safe) using the Hanning taper based on Fourier analysis. A 500 ms sliding window with 50 ms time steps was applied. The power values were extracted from the transformed data in four frequency bands: θ (4–7 Hz), α (8–12 Hz), β (13–30 Hz), and γ (31–40 Hz).

Four regions of interest (ROIs) were defined for the analysis: frontal, parietal, temporal, and occipital (Figure 2). Power was extracted from each ROI in each frequency sub-band during a 0–900 ms window with a 100 ms step length for each participant. Outliers, defined as points outside the three standard deviations from the mean, were detected and subsequently replaced by the nearest element that was not considered an outlier.

## 2.5 Latent profile analysis

Latent profile analysis (LPA) was conducted to divide the participants into subgroups with different hazard recognition performance levels. LPA, which assumes that people can be classified using the degree of probability into categories of different configural profiles of personal attributes, has recently received increasing interest in occupational behavior research (Spurk et al., 2020). In this study, LPA was performed using the *tidyLPA* R package (Rosenberg et al., 2018). The recognition accuracies of the participants for various types of hazards ( Table 1) were used as categorical latent variables. Following the guidelines in existing work (Nylund-Gibson et al., 2007), we adopt a two-step approach. First, consistent with previous work (Gabriel et al., 2015), we determined the optimal number of latent profiles by specifying two latent profiles initially and increasing their number until the model fit did not increase further. The Bayesian information criterion (BIC, recommended by (Nylund-Gibson et al., 2007)) was computed to determine the optimal number of profiles, where a low BIC value implied a good fitting model. Second, LPA posterior probabilities were used to divide the participants into different profile classes (Morin et al., 2001). These categories were named on the basis of the probability of a correct response to any trial for each hazard type.

**Table 1** Descriptions of five hazard types

| Hazard Type | Descriptions of hazards | Example images of hazards | |
|---|---|---|---|
| | | Hazardous | Safe |
| Electric leakage | Overhead power lines, unprotected electrical panels, open electrical compartment, etc. | 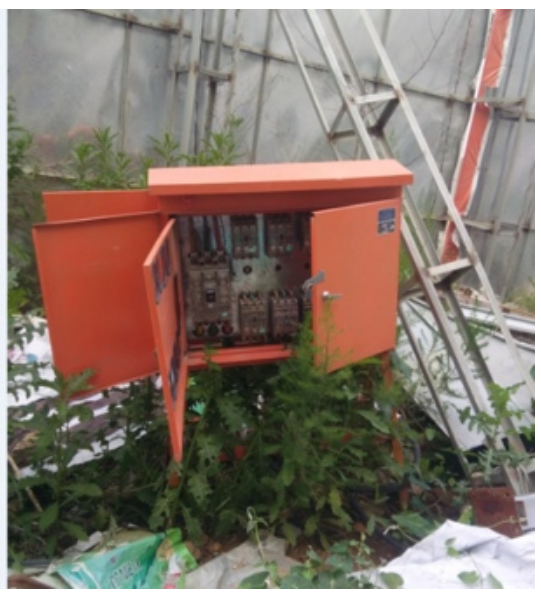 | 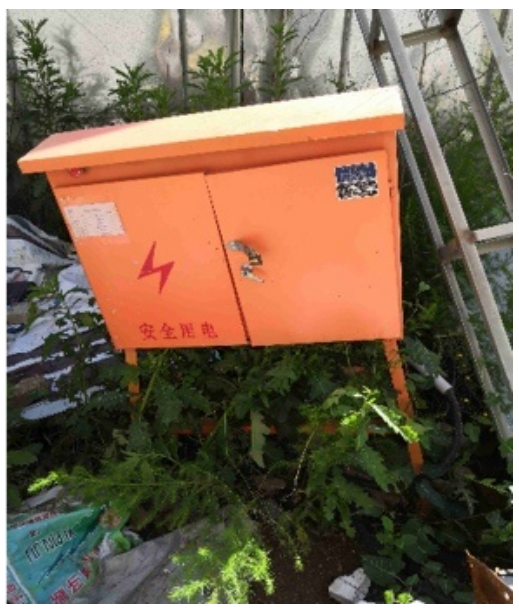 |
| Lack of edge protection | Slips, trips, and falls from heights (elevator hole, stair edges, suspended platform, etc.) without edge protection | 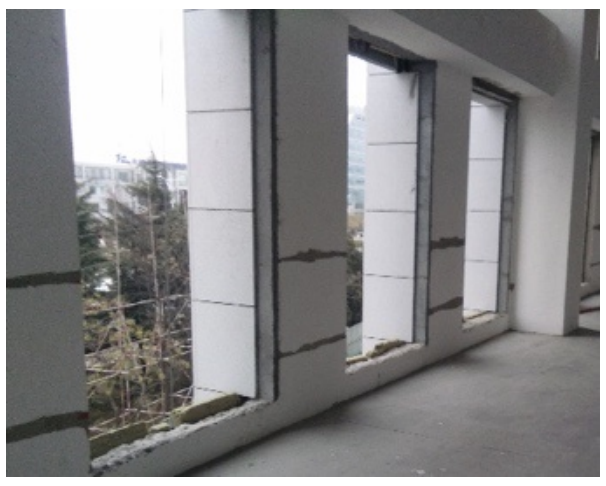 | 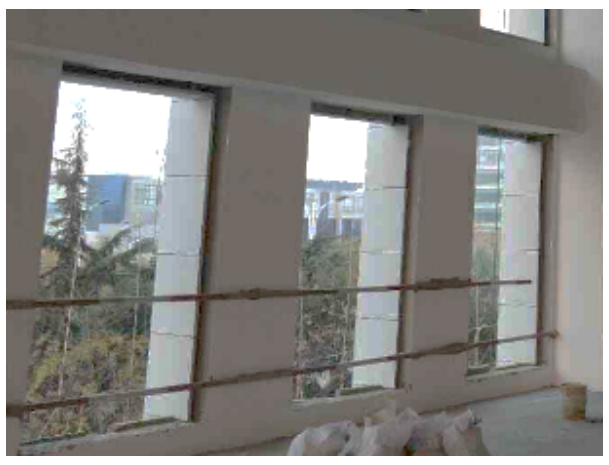 |
| Structural instability | Unstable temporary structures such as scaffolding, formwork, and shoring | 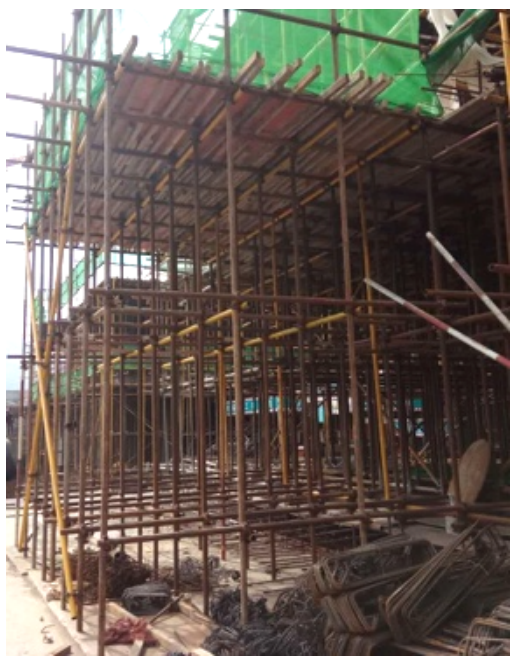 | 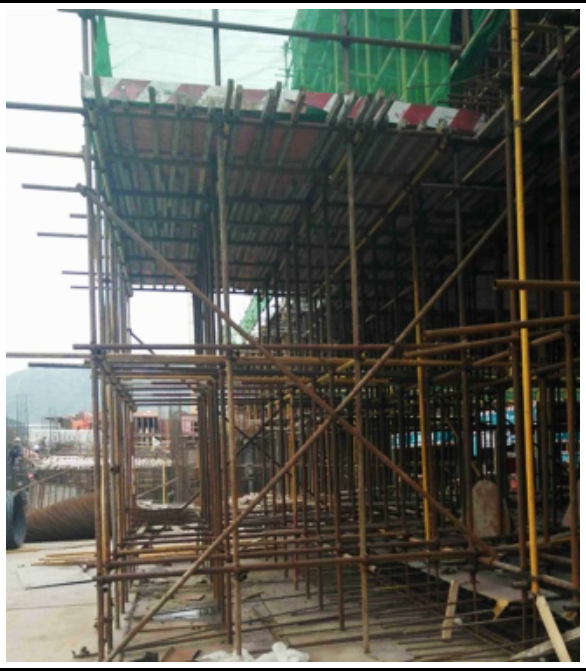 |
| Lack of safety protection equipment | Hard hat/safety belt not used, insufficient ropes for gondola, etc. | 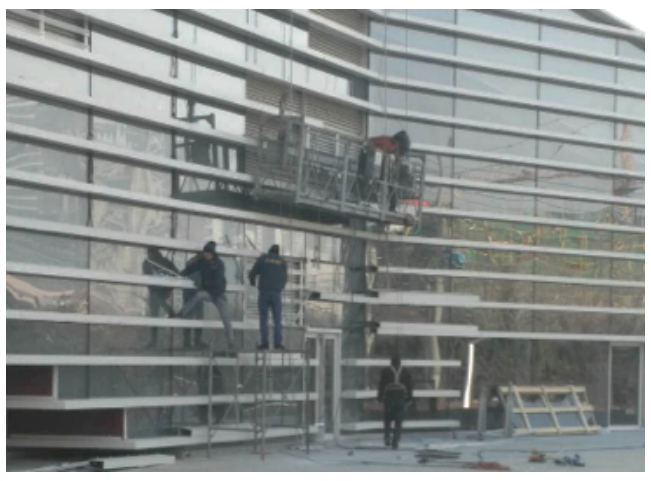 | 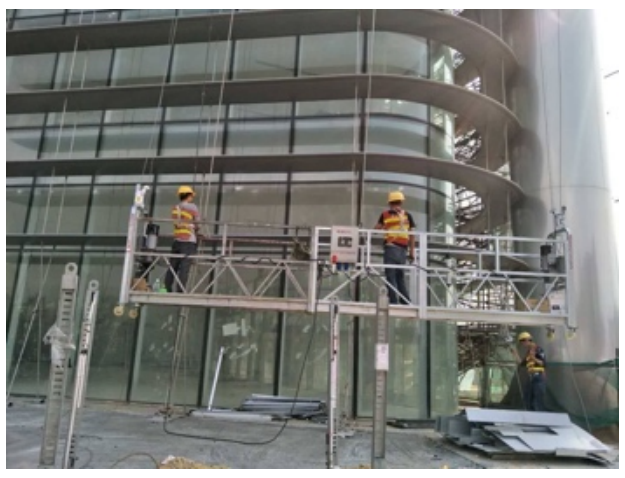 |
| Other | Obstacles on floor can cause tripping, inappropriate storage of chemicals susceptible to fire and explosions, etc. | 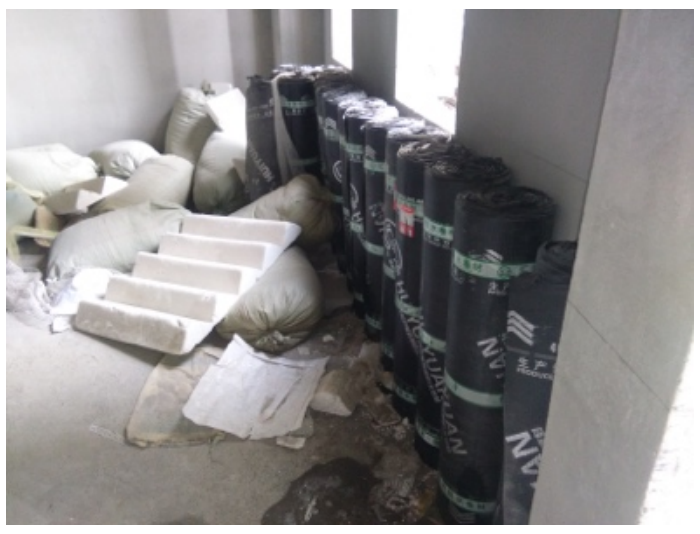 | 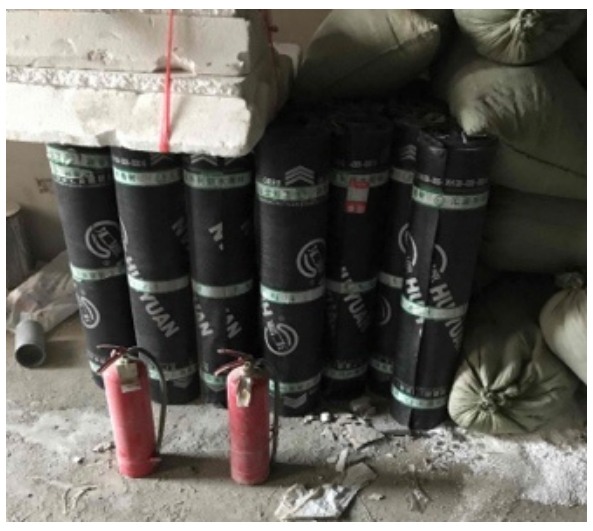 |

### 2.6 Analysis of sensitivity and response bias in signal detection theory

The SDT approach ( Figure 3) was used to analyze the sensitivity (denoted as $d'$ , reflecting the ability to differentiate between hazardous and safe scenarios) of each participant. First, the number of hits (i.e., hazardous images categorized as “hazardous”) and false alarms (i.e., safe images classified as “hazardous”) were calculated. To avoid errors in the following step, extreme values (i.e., 0 and 1) of the hit and false alarm rates were adjusted by replacing them with $0.5/n$ and $(n-0.5)/n$, respectively, where $n$ is the number of trials (Macmillan & Kaplan, 1985). Subsequently, the sensitivity ($d'$ ) and response bias (β) were calculated using Equations (1) and (2), respectively (Wickens, 2000). A large $d'$ value implies that the participants are good at differentiating between hazardous and safe images. β reflects a relative criterion location. Specifically, a β value less than 1 signifies a liberal response bias (i.e., a bias toward responding "yes" or endorsing hazardous images), whereas a β value greater than 1 implies a conservative response bias (i.e., a bias toward a "no" response or endorsing hazardous images as safe). β = 1 implies a neutral point where neither response is favored.

$$d' = Z(\text{Hit}) - Z(\text{False alarm}) \quad (1)$$

$$\beta = \exp((-0.5\ (Z(\text{Hit}) + Z(\text{False alarm})) \times d'), \quad (2)$$

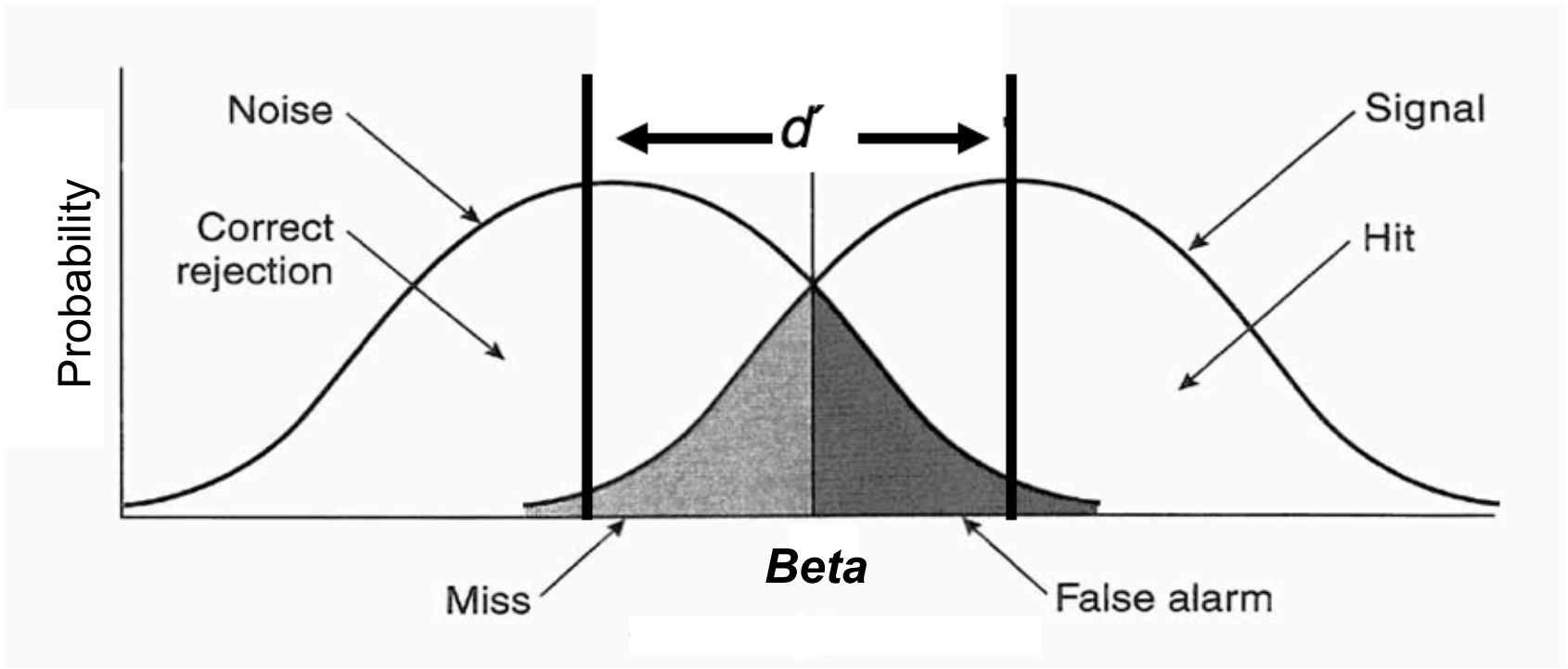

**Figure 3** Graphical representation of SDT model (Wickens, 2000).

## 2.7 Bayesian modeling

The following three steps are involved in the construction of the Bayesian model (Figure 4): (1) a generative model is defined, (2) the inference process of the observer is specified, and (3) the estimated distributions of the observer are computed. The main aim of this study is to investigate whether EEG recordings can be used as an external observation method for internal representations of a world state.

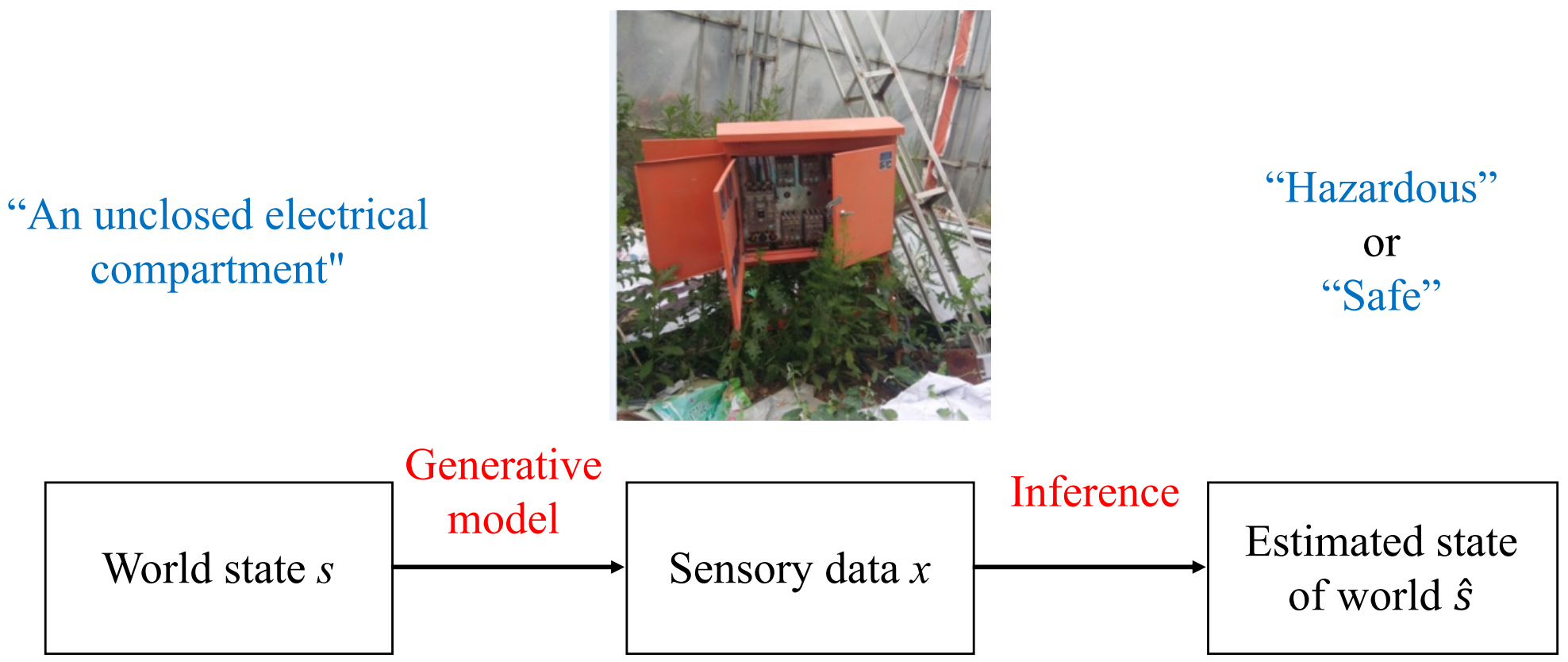

**Figure 4** Schematic of Bayesian model for hazard recognition task. Probability distributions that describe sensory data are generated from the state of the world constituting the generative model. These distributions are specified in step 1 of the Bayesian model. Each participant infers to estimate the world state. This estimate is expressed in step 2 of the Bayesian model. Across many participants, estimation follows a distribution for the given true $s$. This distribution is specified in step 3 of the Bayesian model. Note: $\hat{s}$ might be correct or incorrect.

**Step 1: Generative model is specified**

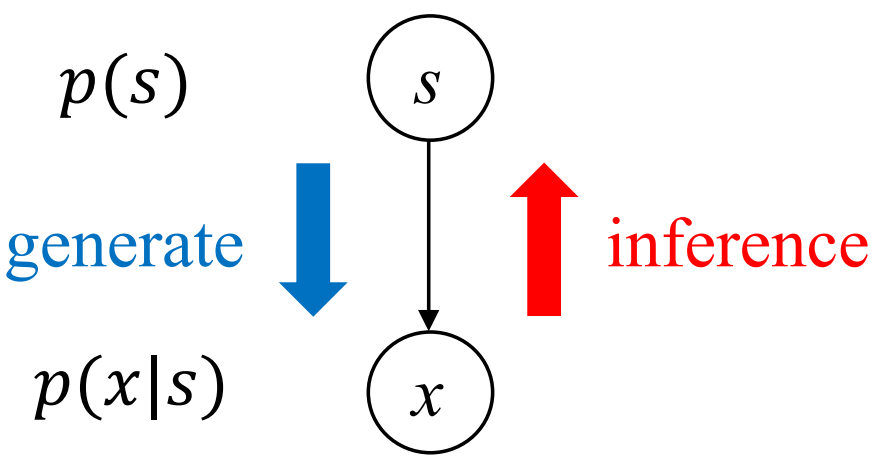

**Figure 5** Graphical representation of generative model. Each node represents random variable. Stimulus (state of the world) $s$ produces noisy measurement $x$. Arrows represent stochastic dependencies between variables.

A generative model describes the probabilistic structure of a task. A physical stimulus induces activity in the nervous system, which is referred to as sensory data or internal representation of the stimulus.

The relationship between these two variables is shown in Figure 5. The noise distribution is the distribution of the measurement $x$ for a given stimulus value $s$, denoted as a conditional distribution $p(x|s)$. Based on SDT, the noise is obtained from a Gaussian distribution. The noise distribution is defined by Equations (3) and (4) and is depicted in Figure 6. As shown in Figure 6(a), a hazardous stimulus (denoted as $s_+$) induces stronger neural activity than a safe one (denoted as $s_-$); however, the variances of the two distributions are identical.

$$p(x|C=1) = \frac{1}{\sqrt{2\pi\sigma^2}} e^{-\frac{(x-s_+)^2}{2\sigma^2}} \quad (3)$$

$$p(x|C=-1) = \frac{1}{\sqrt{2\pi\sigma^2}} e^{-\frac{(x-s_-)^2}{2\sigma^2}}, \quad (4)$$

where $C$ is the stimulus label ($C = 1$ is equivalent to $s = s_+$ and $C = -1$ is equivalent to $s = s_-$).

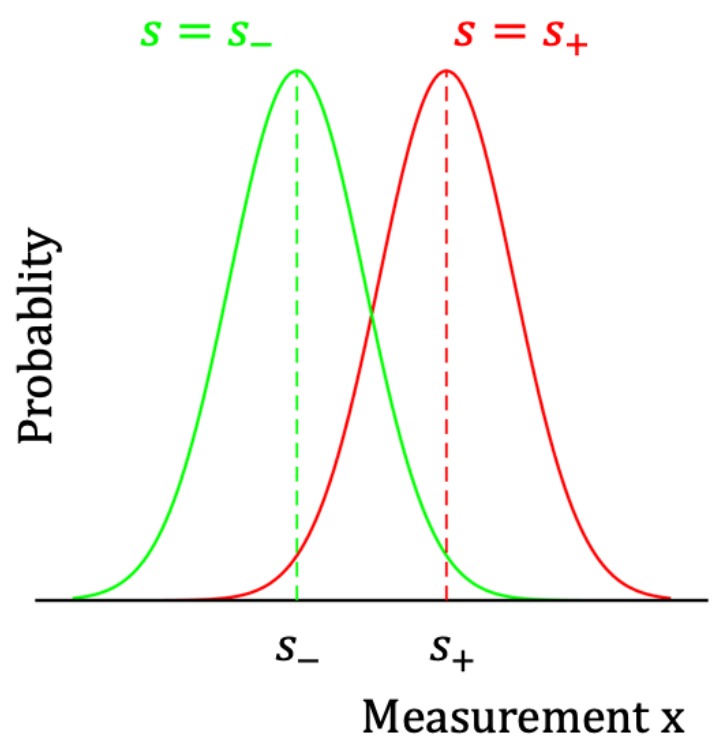

**Figure 6** Gaussian distribution of measurement $x$ for given true stimulus $s$. Conditional distribution $p(x|s)$ is also called noise distribution.

**Step 2: Inference is conducted**

The inference process demonstrates how the observer computes and interprets the posterior distribution, which is hypothesized to be a Bayesian process.

a) Posterior

According to Bayes' theorem (Equation 5), because $p(x)$ does not depend on the "state of the world," the posterior is proportional to the product of the likelihood and the prior.

$$p(s|x) = \frac{p(x|s)p(s)}{p(x)} \text{ , (5)}$$

where $p(x|s)$ denotes the likelihood function of the state of the world and $p(s)$ and $p(s|x)$ denote the prior and posterior distributions over the state of the world, respectively.

A hazard is denoted by the binary variable *C,* which is equal to 1 when a hazard is present and −1 when it is absent. The odds ratio is expressed by Equation (6), and that after the log transformation is given by Equation (7).

$$\frac{p(C=1|x)}{p(C=-1|x)} = \frac{p(x|C=1)\,p(C=1)}{p(x|C=-1)\,p(C=-1)} \quad (6)$$

Log-posterior ratio (log odds ratio):

$$log\frac{p(C=1|x)}{p(C=-1|x)} = log\frac{p(x|C=1)}{p(x|C=-1)} + log\frac{p(C=1)}{p(C=-1)} \quad (7)$$

When the prior is flat, that is, $p(C=1) = p(C=-1) = 0.5$, the optimal decision is based on the log-likelihood ratio, denoted by $d(x)$ (W. Ma et al., 2011) in Equation (8).

$$d(x) = log\frac{p(C=1|x)}{p(C=-1|x)} = log\frac{p(x|C=1)}{p(x|C=-1)} \quad (8)$$

For binary decisions, the maximum-a-posteriori (MAP) is estimated by the sign of the log-posterior ratio. When $d(x)$ is positive, the observer responds as “hazard present” (as expressed in Equation (9)). When the prior is not flat, $d(x)$ is compared to a decision criterion that differs from zero. The absolute value of *d* is a measure of confidence.

$$\hat{C} = \begin{cases} 1, & if\ d(x) > 0 \qquad (9a) \\ -1, & if\ d(x) < 0 \qquad (9b) \end{cases},$$

where $\hat{C}$ denotes the world state reported by the observer.

**Step 3: Distribution of MAP estimates is calculated**

In the third step of the Bayesian model, when *x* is obtained, the probability that Equation (9a) is satisfied must be computed as either $p(x|C=1)$ or $p(x|C=-1)$. As shown in Figure 7, Equation (9a) is satisfied if the measurement falls to the right of the vertical line, when $d(x) > 0$. Thus, graphically, the probability that Equation (9a) is satisfied corresponds to the area under the probability density function to the right of the line.

Because $x$ is a random variable obtained from $\Pr(x|C)$, $\widehat{C}$ is also a random variable. For a binary discrimination task, the distribution of MAP estimates has four discrete probabilities: hit, false alarm, miss, and correct rejection rates. These four rates can be reduced to two because the probability of estimating $\hat{C} = 1$ is 1 minus that of estimating $\hat{C} = -1$.

$$\text{Hit rate} = \Pr\left(\hat{C} = 1 \middle| C = 1\right) = Pr_{x|C=1}(d(x) > 0) \quad (10)$$

$$\text{false alarm rate} = \Pr\left(\hat{C} = 1 \middle| C = -1\right) = Pr_{x|C=1}(d(x) > 0) (11)$$

The probabilities expressed in Equations (10) and (11) were mathematically determined by calculating the area corresponding to the integration of the decision variable density function from zero to infinity. However, this integral is a cumulative density function of the normal distribution; therefore, it cannot be evaluated analytically. A typical alternative is to conduct Monte Carlo simulations. For $C = 1$, 10000 unbiased samples $x$ were generated to estimate the probability of decision variable, $d(x)$.

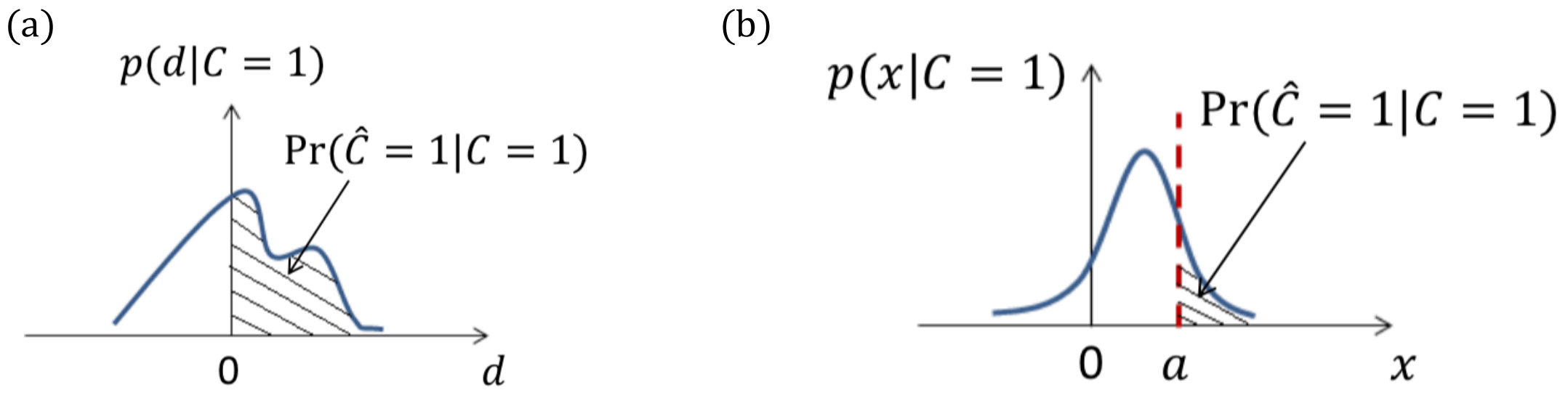

**Figure 7** Class-conditioned distributions of (a) decision variable and (b) measurement.

Because the brain activity measured by EEG is scalar, the decision rule, $d(x) > 0$, can be equivalently expressed as $x > a$ (Figure 7(b)). As the probability of the decision variable can be expressed by likelihood functions, as expressed in Equation (12), the decision rule for reporting C = 1 (i.e., Equation (9a)) can be simplified to Equation (13). A graphical representation of the Bayesian model is shown in Figure 8.

$$d(x) = log\frac{p(x|C = 1)}{p(x|C = -1)} = \frac{s_+ - s_-}{\sigma^2}\left(x - \frac{s_+ + s_-}{2}\right) > 0 \quad (12)$$

$$x > \frac{s_+ - s_-}{2} \quad (13)$$

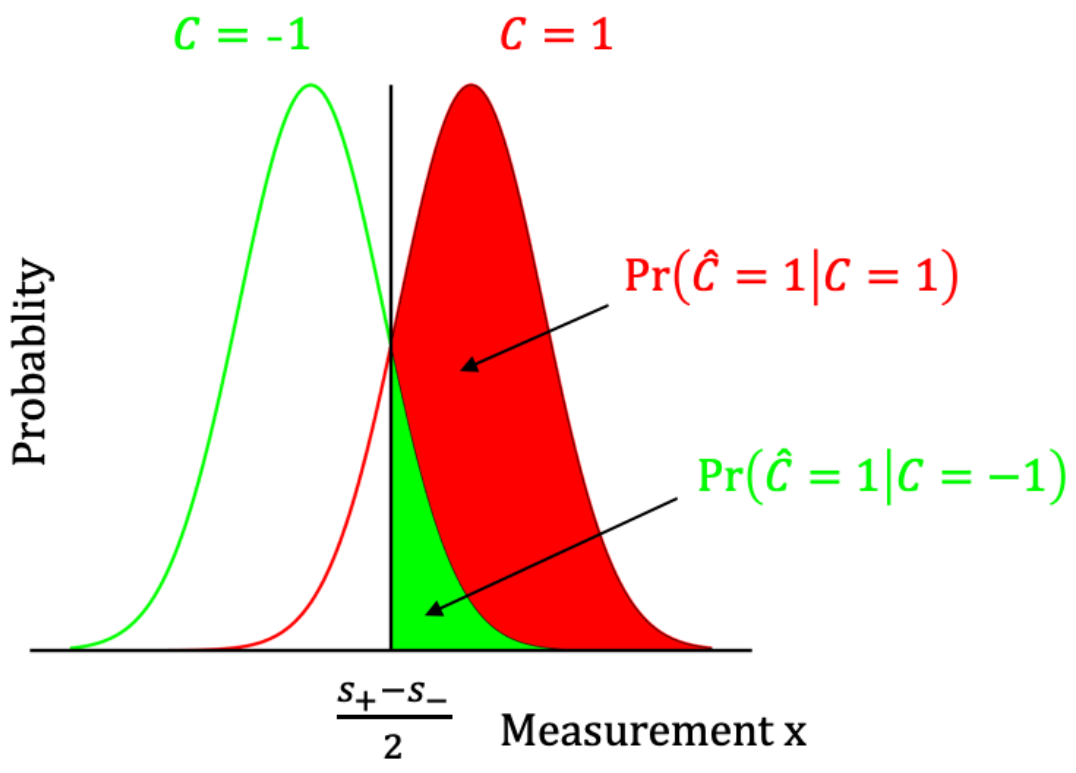

**Figure 8** When Gaussian distributions have identical σ for both stimuli and prior is flat, the decision criterion for reporting $s = s_+$ is $(s_+ - s_-)/2$. The side on which the $(s_+ - s_-)/2$ measurement falls determines the MAP estimate of the observer. The probability that the MAP estimate is *s*+ is equivalent to the shaded area when the true stimulus is *s*- (green) or *s*+ (red).

### 2.7 Receiver operating characteristic curve

Section 3.6 mentions that the generalized hit and false alarm rates of the Bayesian model are equal to the area under the *C* = 1 distribution of the decision variable $\mathrm{p}(d|C = 1)$ or $\mathrm{p}(d|C = -1)$, to the right of a particular criterion *k* (which was previously typically zero). Specifically, hit and false alarm rates can be associated with any criterion that divides the two adjoining decision regions. This corresponds to the decision criterion, *k*, moving continuously along the decision axis and producing hit and false alarm rates at each value. Plotting these hit rates against the false alarm rates produces a smooth curve passing through the origin and (1,1) (Figure 9).

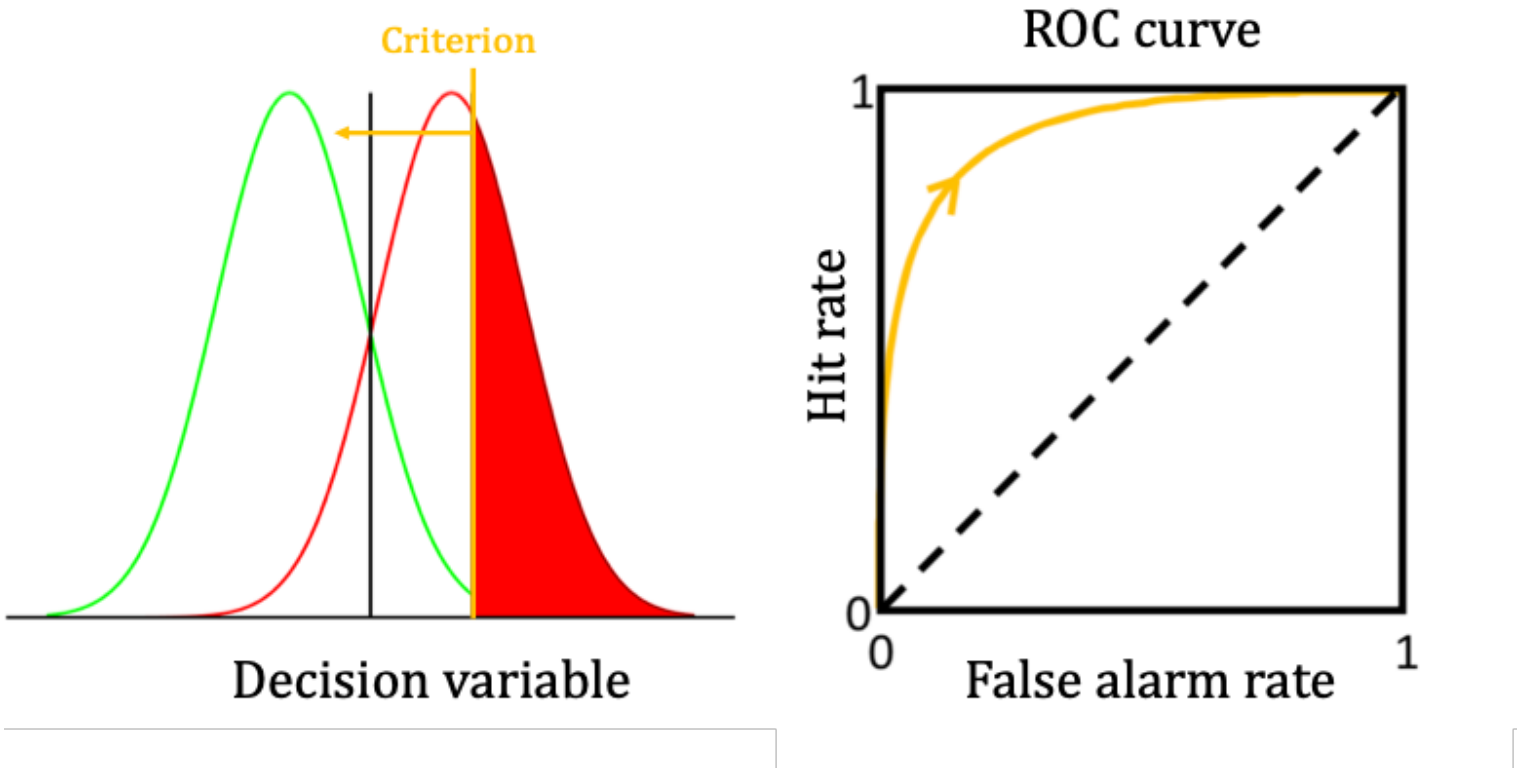

**Figure 9** Distribution of decision variable and receiver operating characteristics.

## 2.8 Numerical simulations of decision criteria for confidence ratings

In a binary decision, the sign of the log-posterior ratio determines the MAP decision. This ratio also has a magnitude or absolute value, where a small absolute value implies that the posterior probabilities of the two alternatives are close to each other. Therefore, a possible measure of confidence in a binary decision is the magnitude of the log-posterior ratio. Because this confidence corresponds to the distance from the origin on the decision variable axis, the Bayesian model can be used to predict the responses of the observer to a discrimination task and determine the confidence level associated with this decision (Figure 10(a)). Notably, the critical challenge is to estimate the dividing line between low- and high-confidence levels. To develop a practical BCI application, we provided confidence ratings for low-, medium-, and high-confidence levels. Thus, six possible responses existed: two category estimates and three confidence ratings. The observer can choose one of these by determining the value of the decision variable in which the six decision regions fall (Figure 10(b)). These regions were separated using five decision criteria.

An α-band power of 200–0 ms prior to the response was used to estimate the decision confidence rating.

A receiver operating characteristic (ROC) curve was obtained by fitting the points from the response frequencies to each of the six response categories for each of the two classes using cubic spline interpolation. Numerical simulations were conducted to identify a set of decision criteria with the best fit to the ROC curve, as described in Section 3.7 (Figure 18d). Curve matching was assessed using Euclidean distance.

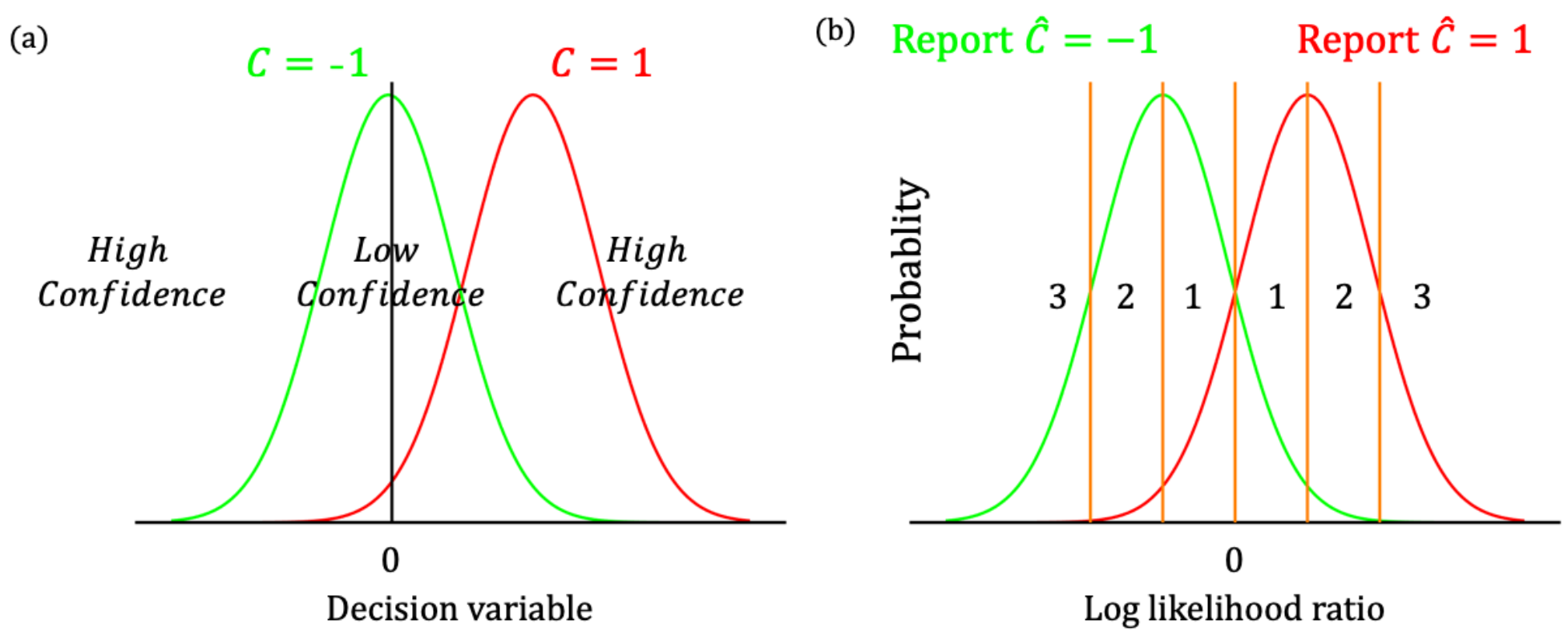

**Figure 10**(a) Absolute value of decision variable is a measure of confidence. (b) Response categories in hypothetical measurement with confidence ratings.

## 3. Results

### 3.1 Feasibility of EEG measurements to observe internal representations to hazard recognition tasks

LPA was used to divide the participants into two subgroups because this categorization presented the lowest BIC (Figure 12). These subgroups are “good performers” and “bad performers,” where the first subgroup is likely to perform well in hazard recognition tasks, whereas the latter has a more ambiguous probability of performance accuracy (Figure 11). The posterior probabilities showed a 74.3% (“good performers”)–25.7% (“bad performers”) split among the participants.

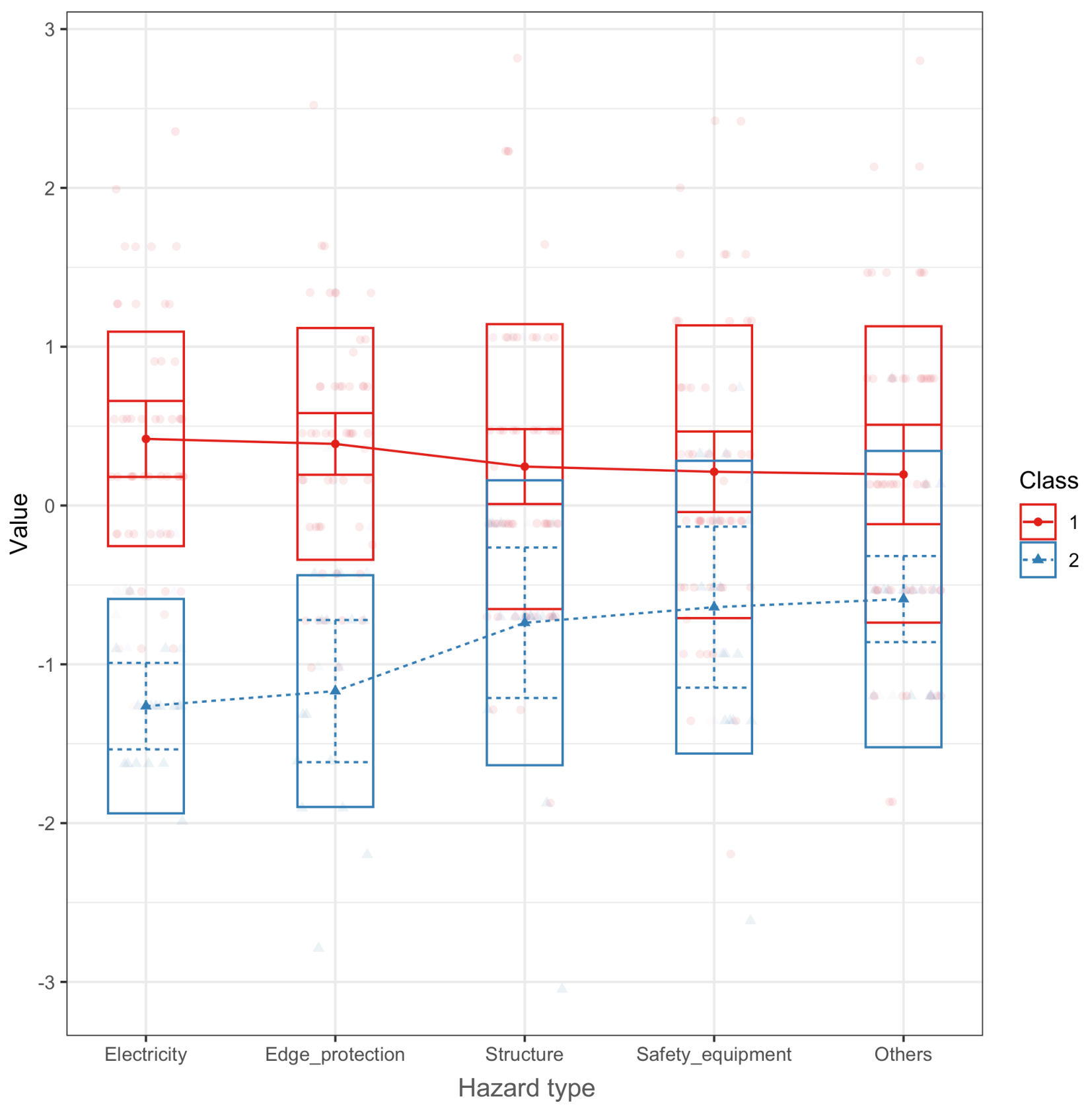

**Figure 11** Two profile classes of hazard recognition performance identified among participants: (1) good performers and (2) bad performers. Column centering and scaling are conducted to demonstrate the accuracy performance separation of each profile category. Centering involves subtracting column means from their corresponding columns, whereas scaling is conducted by dividing columns by their standard deviations.

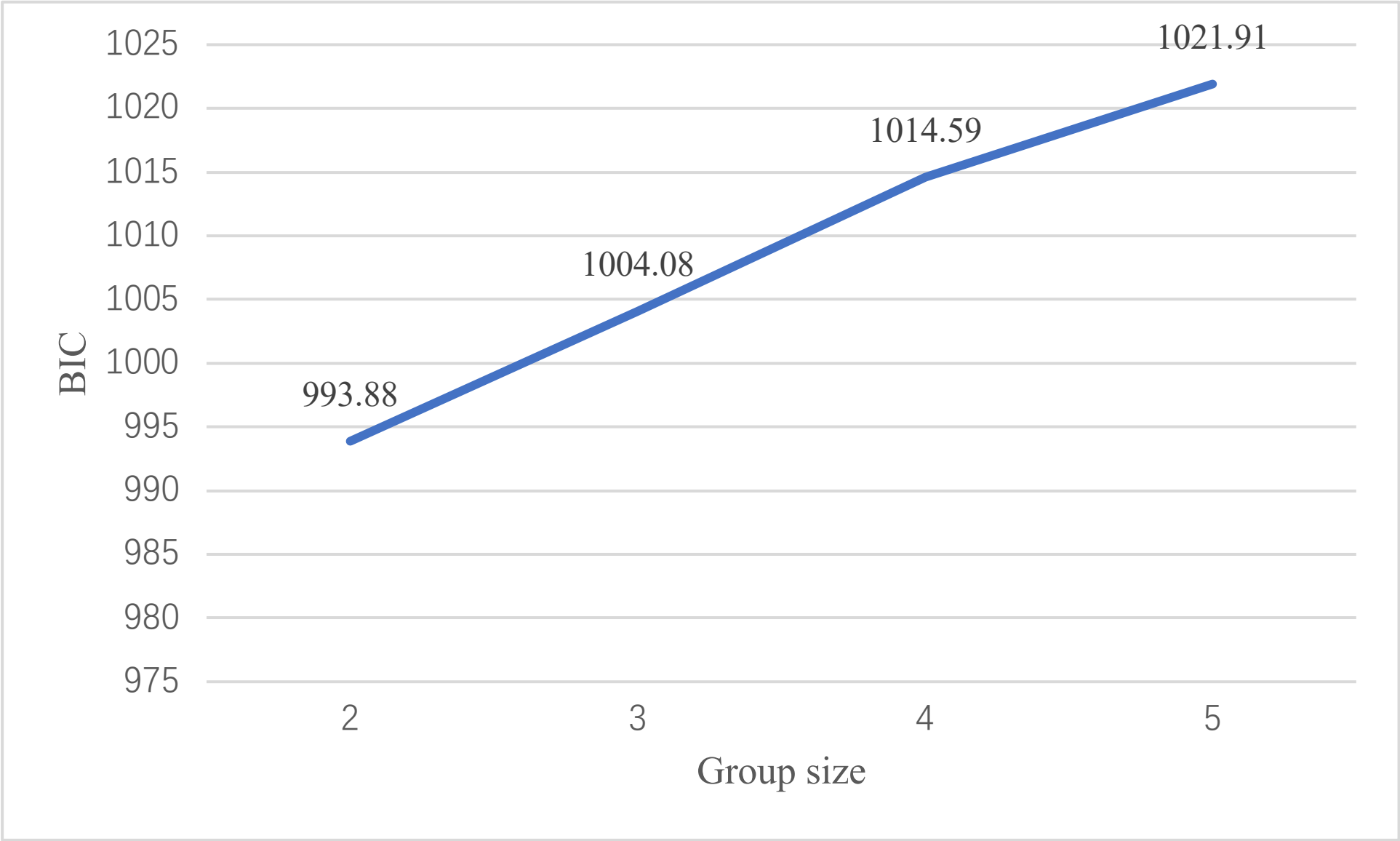

**Figure 12** BICs for multiple group sizes in LPA.

Because the powers induced by the hazardous and safe stimuli did not fit normal distributions (Lilliefors test, $p < 0.05$), a Wilcoxon signed-rank test was performed to compare the power induced by hazardous stimuli with that induced by safe stimuli for each time segment from each electrode in the four frequency subbands in each subgroup. In the "good performers" subgroup, a significantly strong activation was produced by the hazardous stimuli in the θ, α, β, and γ bands from the multichannels 400 ms after the stimulus onset (Figures 13(a) and 14). However, in the "bad performer" subgroup, the four frequency sub-band powers scarcely showed a significant difference between the two conditions (Figure 13(b)).

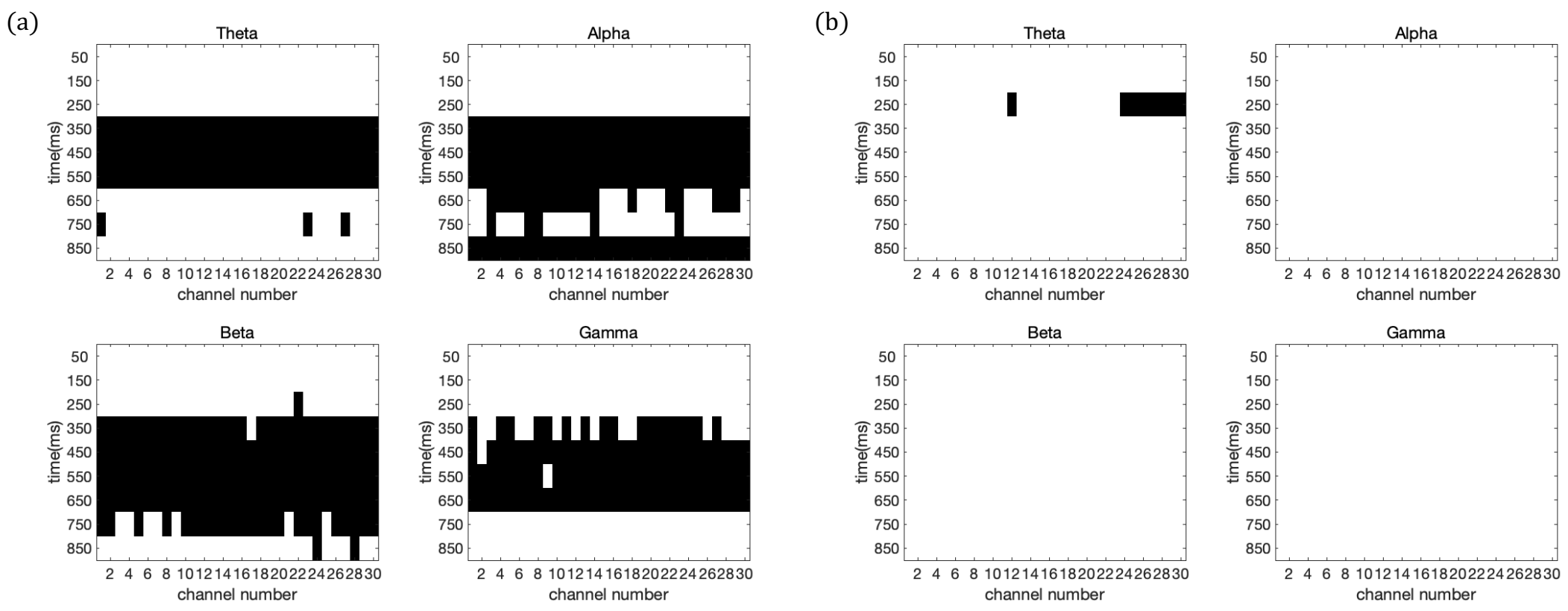

**Figure 13** Results comparing EEG powers induced by two conditions (hazardous and safe) in the (a) "good performers" and (b) "bad performers" subgroups for each time segment (0–100, 100–200, 200–300, 300–400, 400–500, 500–600, 600–700, 700–800, and 800–900 ms) from each channel (Fp1, Fp2, Fz, F3, F4, FC1, FC2, FC5, FC6, Cz, C3, C4, T3, T4, CP1, CP2, CP5, CP6, Pz, P3, P4, P7, P8, PO3, PO4, PO7, PO8, Oz, O1, and O2). Blocks filled in black denote corresponding $p < 0.05$.

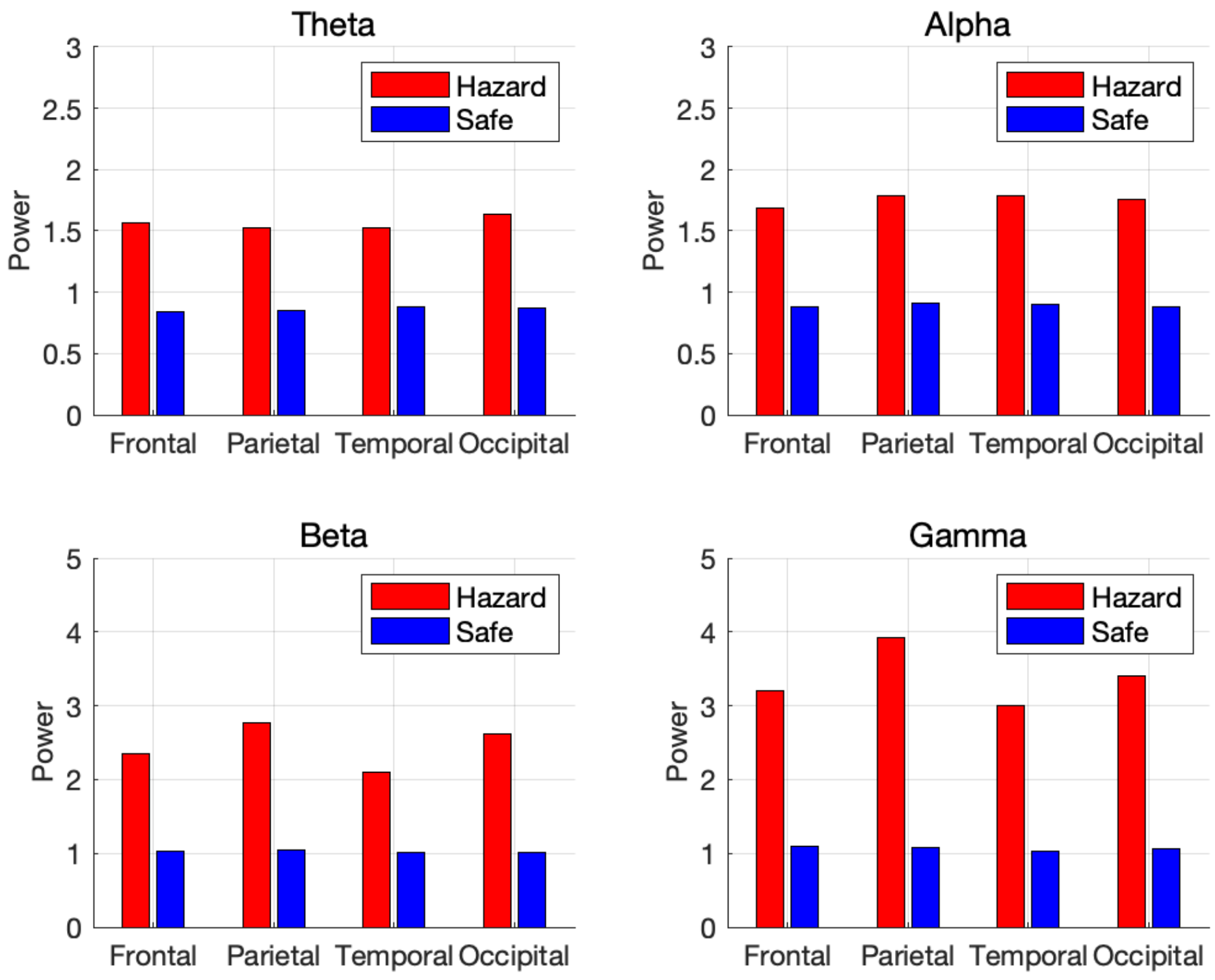

**Figure 14** EEG power induced by two conditions for four ROIs in the "good performers" subgroup (400–700 ms).

The difference in power values was calculated by subtracting the power induced by the safe stimuli from that induced by the hazardous stimuli for each participant. The correlation between the power difference and sensitivity measure (d′) was examined for all participants. Because the d′ data did not fit a normal distribution (Lilliefors test, $p < 0.05$), Spearman's correlation was conducted. Significant correlations were found in the θ, α, and β bands 200–500 ms after the stimulus onset ($p < 0.05$, $r \approx 0.25$; Figure 15).

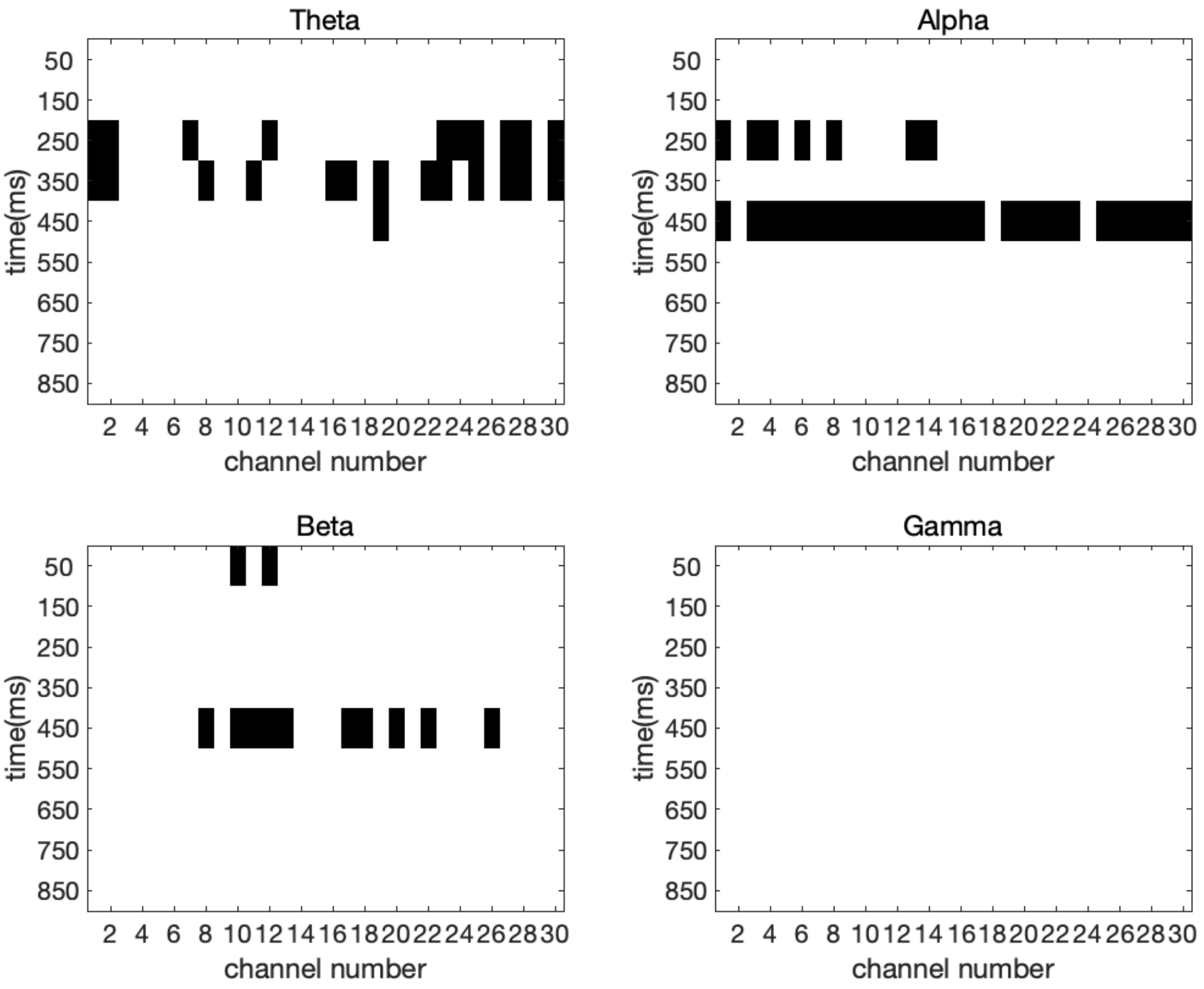

**Figure 15** Correlation between EEG power difference and sensitivity measure (d′) for all participants in each time segment from each channel. Blocks filled in black denote corresponding $p < 0.05$.

### 3.2 Hazard decision threshold characterized by EEG power

A group-level analysis was conducted to estimate the decision threshold characterized by the EEG power above which the participants responded as "hazardous" (Figure 16). The distributions of EEG power data under hazardous and safe conditions are shown in Figure 17(a). Considering that the figure indicates a positive skew, a square root transformation was conducted, and the results are shown in Figure 17(b). Distribution normality was assessed using the Lilliefors test, which revealed that both categories were normally distributed ($P > 0.05$). Monte Carlo simulations were conducted to generate 10000 trials to obtain Gaussian convergence using the mean and standard variances derived from the transformed data (Figure 17c). The decision variable is computed using Equation (8), and its distribution is shown in Figure 17(d). Using Equation (13), the hazard perceptual threshold of the EEG power can be calculated as:

$$x > \frac{mean(\sqrt{C=1}) + mean(\sqrt{C=-1})}{2} \quad (14)$$

The estimated hit and false alarm rates determined using the MAP distribution (probabilities predicted using Equations (10) and (11)) are listed in Table 2. The α-band EEG power from the prefrontal area 500–700 ms after the stimulus onset showed the minimum hit rate MAP estimation error. In comparison, the empirical behavioral data yielded hit and false rates of 75.97% and 51.45%, respectively.

In addition, the MAP estimates were tested at the individual level. The transformed EEG data of the 38 participants met the assumptions (normality and homogeneity of variance). The average estimated hit and false alarm rates were 56.55% and 43.48%, respectively, which were lower than those from the empirical behavioral data (average empirical hit rate, 72.15%; Wilcoxon signed-rank test, $p < 0.001$; average empirical false alarm rate, 46.97%). Based on Figure 8, the hazard decision threshold characterized by the EEG power was higher than the empirical decision threshold of the participants, resulting in a low MAP estimate. This is consistent with the response bias (average $\beta = 0.81 < 1$), which also implies a liberal response bias.

Because the distribution of the MAP estimates represented only one point on the ROC for the binary discrimination task, it was insufficient to characterize the internal representation model of the participants. Assuming that humans adopt a decision variable and use the optimal criterion for decisions, the distributions of the decision variables calculated from multiple temporal-spatial sources and the corresponding ROC curves parameterized by the decision criterion are shown in Figure 18. These can be used to characterize an individual's internal model.

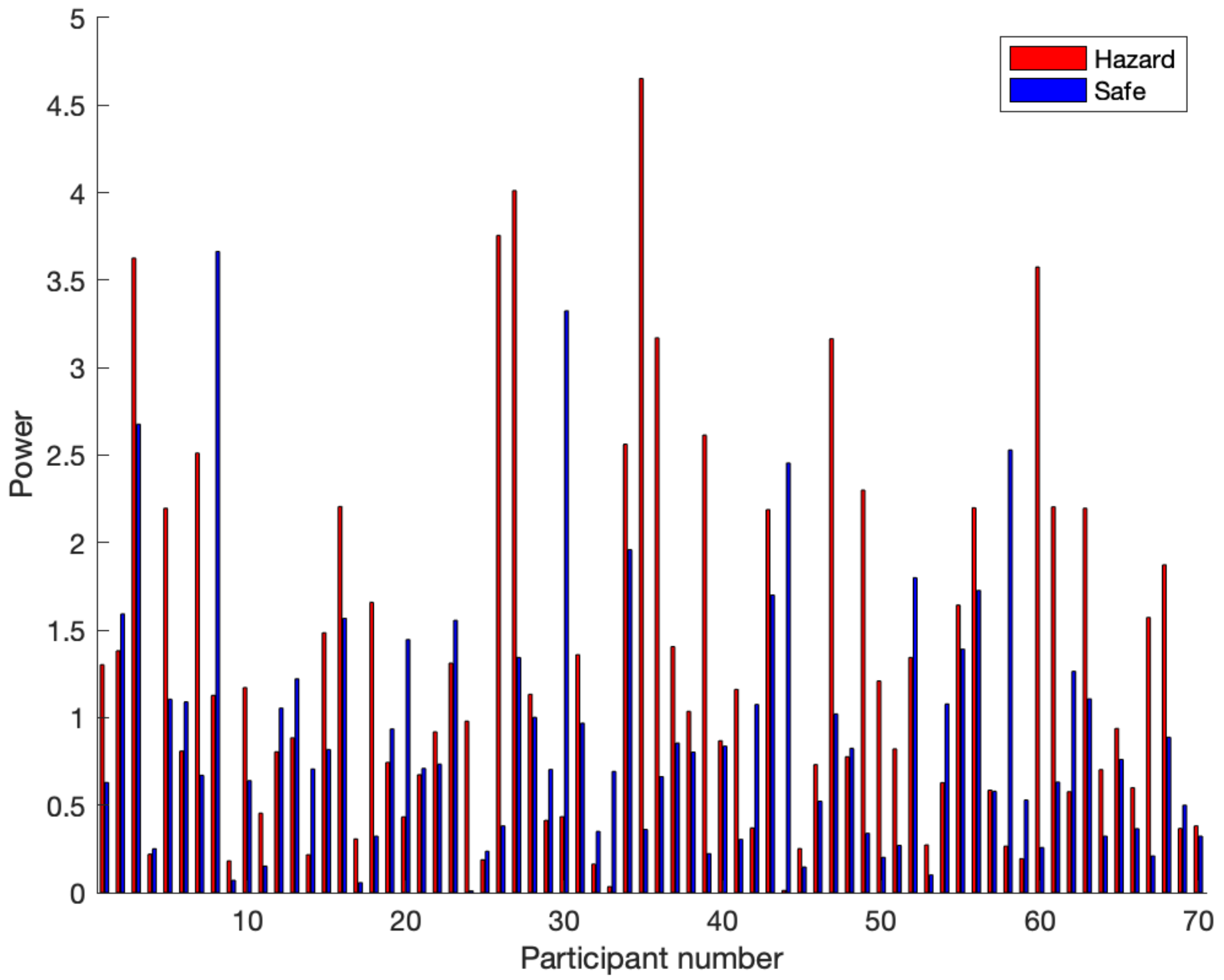

**Figure 16** Bar plot of prefrontal α-band power 500–700 ms after stimulus onset induced by hazardous or safe stimuli for each participant.

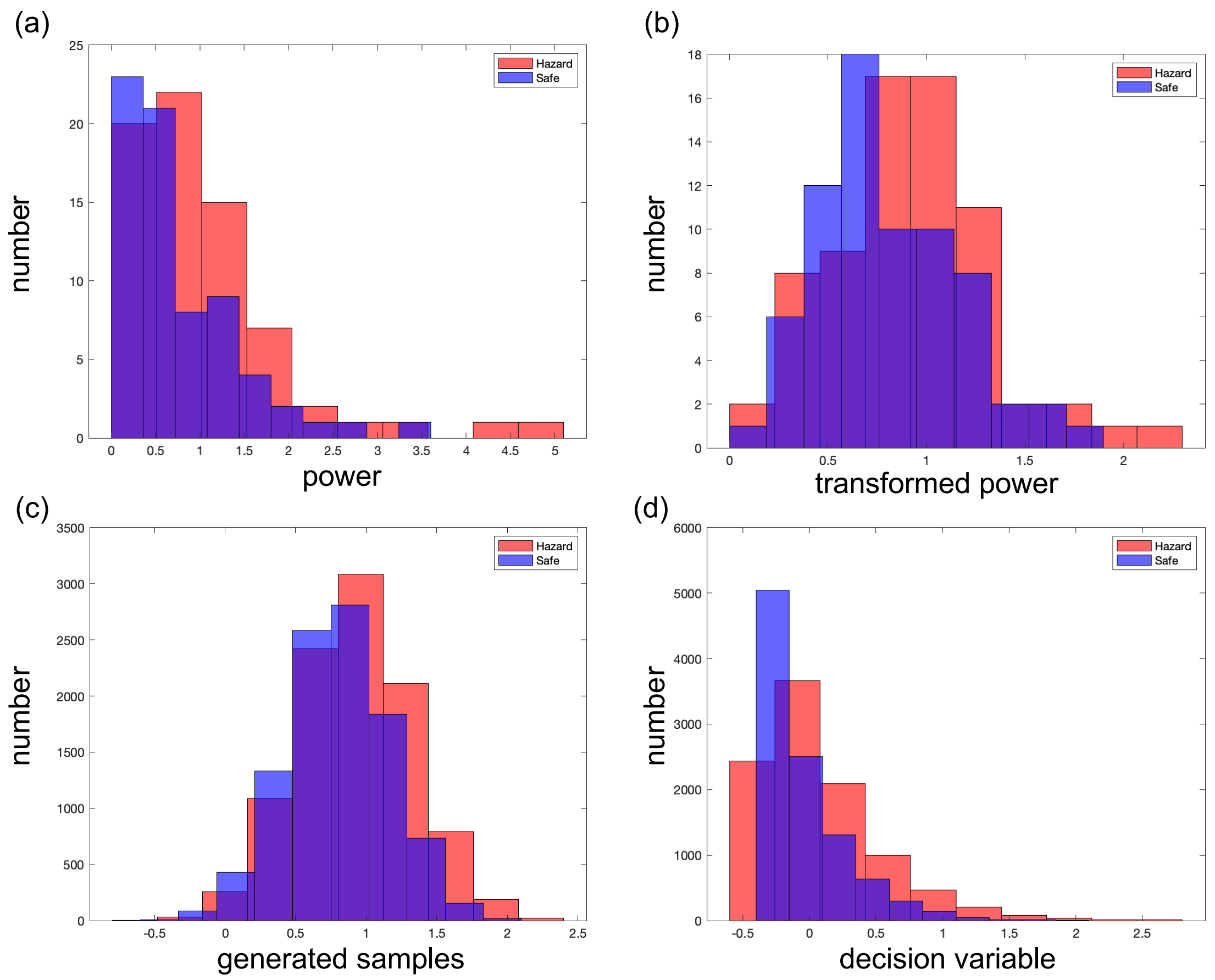

**Figure 17** Distributions of (a) raw EEG power (parietal α-band power 700–800 ms after stimulus onset), (b) EEG power after square root transformation, (c) 10000 samples generated by Monte Carlo simulations, and (d) decision variables calculated using Equation (8).

**Table 2** Estimated hit and false alarm rates from EEG power using MAP distribution and area under curve (AUC) of corresponding ROCs from multiple sources

| Source / Estimated rate | 0–100 ms all | 300–400 ms occipital | 500–700 ms prefrontal | 700–800 ms parietal |
|---|---|---|---|---|
| Hit rate estimated using Equation (12) | 53.01% | 56.66% | 60.94% | 56.98% |
| Hit rate estimated using Equation (13) | 55.94% | 57.88% | 62.84% | 56.85% |
| False alarm rate estimated using Equation (12) | 47.67% | 42.88% | 36.76% | 43.10% |
| False alarm rate estimated using Equation (13) | 51.84% | 48.00% | 42.48% | 44.10% |
| AUC | 0.536 | 0.593 | 0.653 | 0.598 |

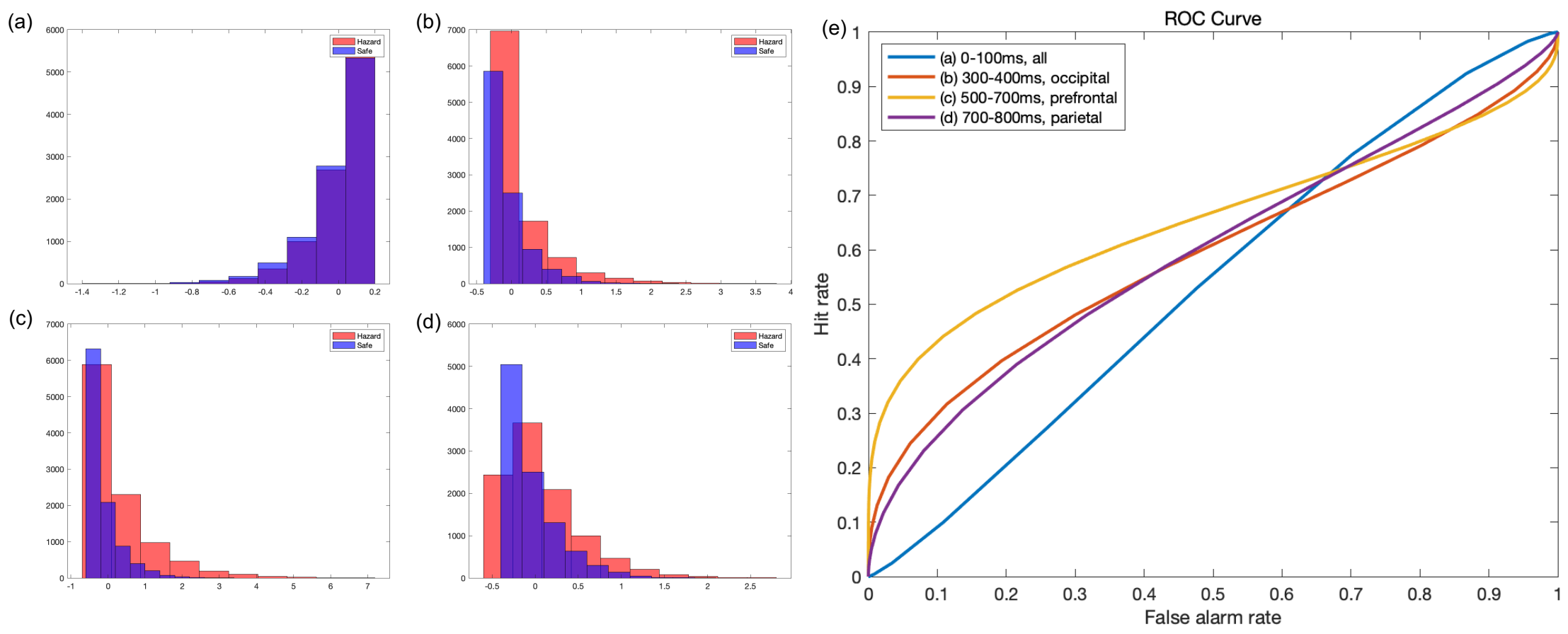

**Figure 18** Distributions of decision variables calculated from EEG power in (a) all electrodes 0–100 ms, (b) occipital area 300–400 ms, (c) prefrontal area 500–700 ms, and (d) parietal area 700–800 ms after stimulus onset. (e) Corresponding ROCs.

### 3.3 Decision criteria for confidence ratings by numeric stimulations

A two-row × six-column table was created to estimate the ROC from the response-locked EEG power (Figure 19(a)). The top and bottom rows correspond to true classes with $C = 1$ and $C = -1$, respectively. Each column corresponds to a response category. The three left and right columns correspond to the $\hat{C} = -1$ and $\hat{C} = 1$ responses, in the order of decreasing and increasing confidence, respectively. Each cell in the table contains the frequency of responses for each category divided by the total number of responses across all categories for that class. Thus, the sum of the numbers in each row was equal to one. Subsequently, a new table was created by cumulatively summing the numbers in the original table from right to left separately for each row (Figure 19(b)). Each column corresponds to a hit-and-false alarm rate pair. The leftmost pair is always (1,1). Finally, the hit rate is plotted against the false alarm rate, as shown in Figure 19(c), and the optimal decision criteria for confidence rating differentiation are summarized in Figure 20(a).

(a)

| True C | Reported C and confidence rating: -1 high | -1 medium | -1 low | 1 low | 1 medium | 1 high | |
|---|---|---|---|---|---|---|---|
| 1 | 0.0797 | 0.2536 | 0.1667 | 0.2826 | 0.1159 | 0.1014 | 1 |
| -1 | 0.1136 | 0.1439 | 0.2803 | 0.2197 | 0.1515 | 0.1288 | 1 |

(b)

| True C | Reported C and confidence rating: -1 high | -1 medium | -1 low | 1 low | 1 medium | 1 high |
|---|---|---|---|---|---|---|
| 1 (Cumulative Sum) | 1.0000 | 0.9203 | 0.6667 | 0.5000 | 0.2174 | 0.1014 |
| -1 (Cumulative Sum) | 1.0379 | 0.9242 | 0.7803 | 0.5000 | 0.2803 | 0.1288 |

(c)

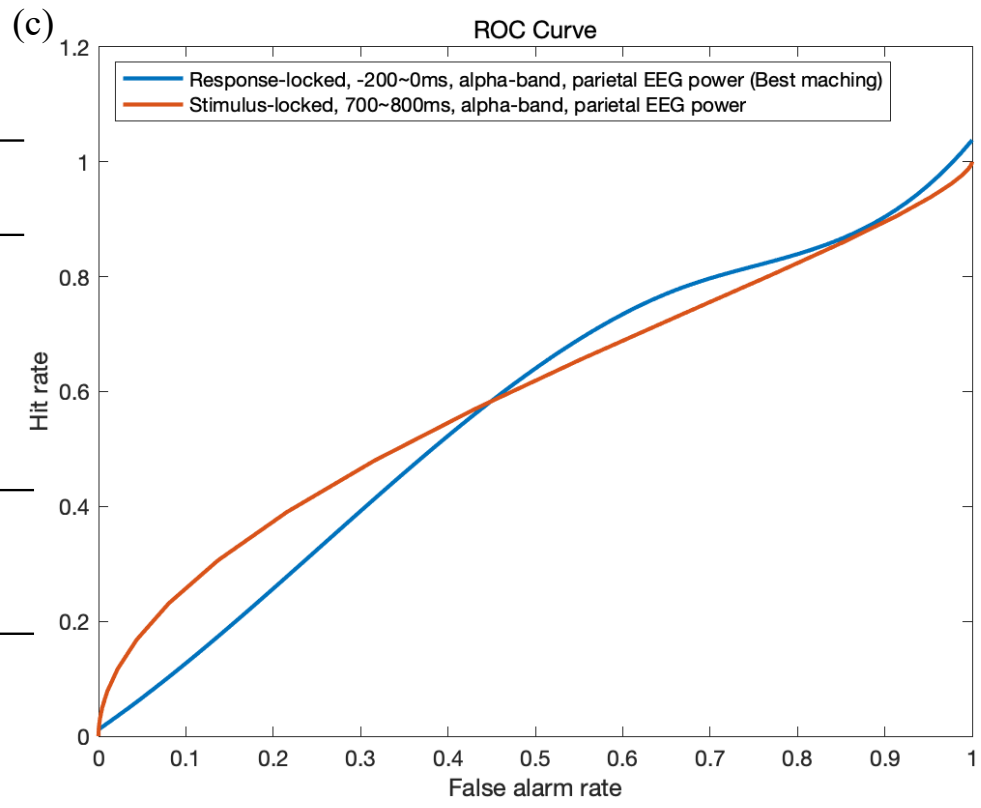

**Figure 19** (a) Response frequencies and (b) cumulative sum of response frequencies in each category (two class estimates × three confidence ratings for each condition). (c) Best matching ROCs estimated from stimulus-locked and response-locked EEG powers.

(a)

| True C | Reported C and confidence rating: -1 high-medium | -1 medium-low | 1 low-medium | 1 medium-high |
|---|---|---|---|---|
| 1 | 3.0 | 1.5 | 1.2 | 3.0 |
| -1 | 3.5 | 1.0 | 1.5 | 3.5 |

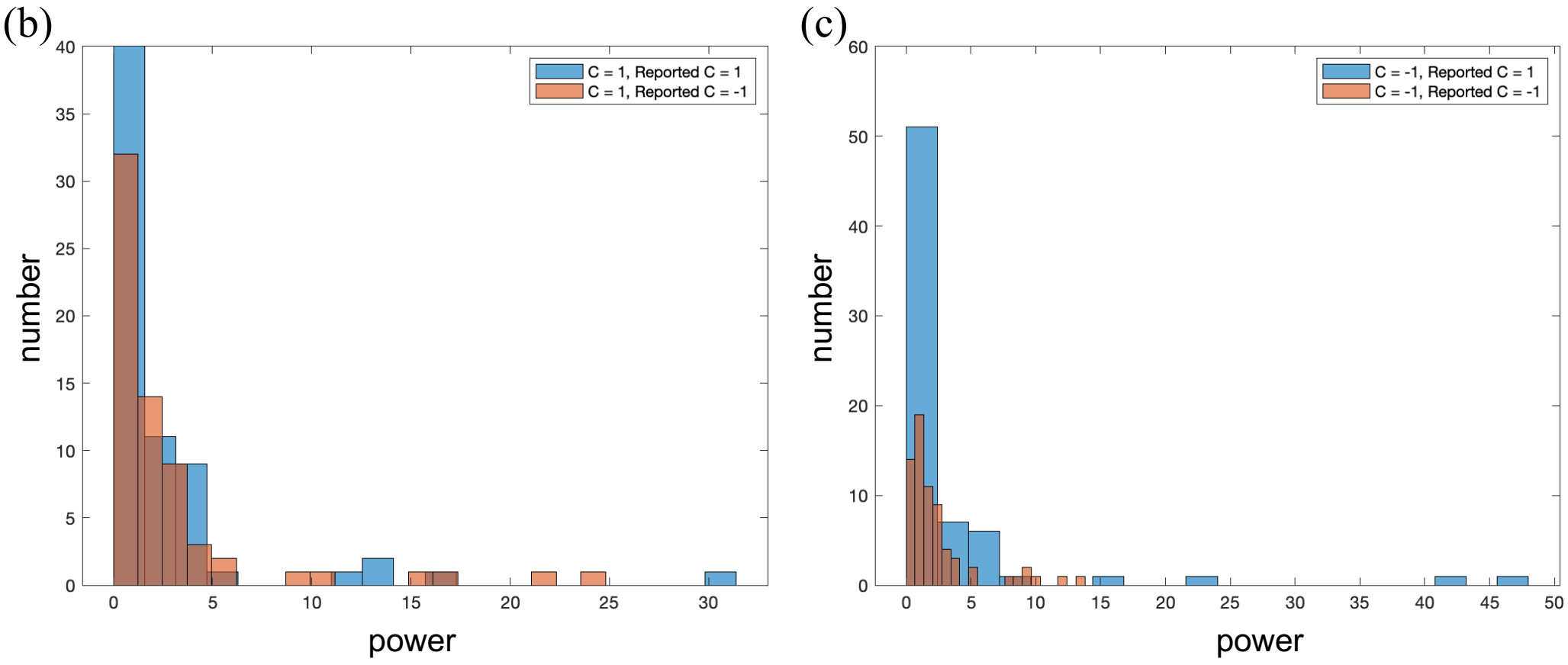

**Figure 20** (a) Optimal decision criteria characterized by parietal α-band EEG power 200–0 ms prior to response for confidence rating differentiation. Distributions of EEG power induced by (b) hazardous and (c) safe stimuli in each response category.

# 4 Discussion

A confidence-weighted method has been proven to be superior to other strategies for integrating decisions from construction workers and automation systems when implementing human-machine collaboration for hazard inspection (Meyen et al., 2021; Poli et al., 2014). Computer vision algorithms typically produce outputs supplemented by a degree of confidence. However, for humans, self-reported confidence weighting is often inaccurate and it requires additional behavior and communication efforts, in addition to from simple communication information for individual decisions. To address this problem, BCI is applied, aiming to provide an intuitive and objective channel of communication for hazard perception and decision-making processes. Specifically, we developed a novel approach for predicting decision choices and estimating decision confidence during hazard recognition by using EEG-measured brain wavelets.

In this study, SDT was used to evaluate the ability of an individual to discern between an information-bearing stimulus (a "hazard") and noise (Albert et al., 2017). The sensitivity (d′) of SDT indicates the distance between the means of the signal and noise distributions, providing an objective estimation of the ability to differentiate between the two stimulus conditions. The power difference (i.e., the difference in brain responses induced by two different stimuli) has been computed in previous neuropsychological studies to reflect the neural computation mechanism for categorization tasks (Heekeren et al., 2004; Shadlen et al., 1996). Because SDT curves represent hypothetical brain activity (Wickens, 2000), we considered that the sensitivity measure and power difference would correspond. Based on the experimental results of a group-level analysis, the participants with superior hazard recognition performance showed significantly larger neural activation across the entire brain for hazardous stimuli, as compared with that in the case of safe stimuli. Moreover, the individual-level analysis revealed a significant correlation between the d′ value and power difference. These findings support our hypothesis that EEG-measured cortical activity can be used to observe internal representations in hazard recognition tasks. Furthermore, the hazard perceptual threshold and hit and false alarm rates can be determined by computing MAP distributions. Thus, we can predict the EEG-equipped worker's responses to hazards by observing whether his brain activation exceeds the perceptual threshold for reporting hazards. We can also predict each worker's hazard recognition performance that is attributed to the sensitivity and identity of vulnerable workers to provide targeted warning or support, which is a significant way to reduce accidents at construction sites.

Response-locked parietal α-band oscillatory power was used as a proxy for decision confidence, which is consistent with previous studies showing that the α-band plays a critical role in attentional modulation (Foxe & Snyder, 2011). The optimal decision criteria, characterized by EEG power averaged over 200–0 ms prior to the response, were identified for high-, medium-, and low-confidence level differentiations. These criteria are critical for practical BCI application of weighing votes, considering the decision confidence for an optimal group decision-making result. The levels are divided according to an optimal Bayesian observer (W. Ma et al., 2011), and by observing the interval within which the EEG-equipped worker's brain activation lies, we can integrate the worker's hazard choice prediction with its uncertainty. This will comprehensively evaluate the worker's vulnerability (associated with his ability to differentiate between hazards) and weigh votes for the worker's hazard choice in situations where hazards are identified by peers or automated agents. This supports the hazard confirmation mechanisms embedded within existing hazard warning platforms and thereby promotes project safety performance.

Furthermore, a key finding is that the formulation of a high-confidence decision, irrespective of its correctness, exerts more EEG power under safe conditions than under hazardous conditions, indicating that more effort is required to differentiate between safety and hazards. This was supported by the underlying internal model for hazard recognition, that is, the evidence accumulation process in Bayesian inference. Under safe conditions, participants assumed a higher perceptual threshold for evidence accumulation to reach decisions than under hazardous conditions. In these situations, the participants demonstrated a conservative response bias, which would further lead to a lower response accuracy under safe conditions compared with that under hazardous conditions (48.55% correct rejection rate compared with a 75.97% hit rate). This finding has important implications for designing safety training programs, because currently, safety training is primarily aimed at improving workers' discernment skills for hazards by equipping them with appropriate internal hazard representations. However, when workers discriminate potentially hazardous scenes, they tend to exert more effort to identify actual safe scenes and are more likely to misclassify them as hazards. Although caution is encouraged, future safety training is expected to help calibrate workers' subjective definition of what sensory cues count as a "hazard" to an appropriate level to avoid unnecessary efforts and misclassification.

This study had several limitations that should be addressed in future studies. First, this study was conducted in a laboratory setting. Although the stimuli used in the experiment were pictures depicting actual construction scenes, considering the dynamics and clutter on actual construction sites, unconsidered factors such as enhanced selective attention and occupied attention capacity may affect cognitive functions and brain activities during hazard recognition. Therefore, future work can employ wearable EEG-based BCIs to collect psychological signals of construction workers on actual construction sites to improve the external validity of the results. Second, this study required participants to perform binary discrimination tasks in potentially hazardous environments; however, in practice, if workers identify hazards, they may respond in various ways, ranging from immediately ratifying the hazardous situation to discarding it. Therefore, further efforts are encouraged to decode workers' intentions and response preferences from their brain activities using a multiple-choice experimental design. Finally, although this study implemented the developed method at an individual level to validate the hazard perceptual threshold determined from EEG recordings, in the future, it is expected to further explore how personal traits may influence the perceptual decision-making processes during hazard recognition based on the developed method.

## 6 Conclusion

This study established a promising novel approach for predicting the response choice of an individual and quantifying the associated confidence level of the choice when identifying hazards. First, we developed a conceptual model under the Bayesian inference framework for a hazard recognition task using hypothetical brain activities to signify the internal representations of humans when discriminating between hazardous and safe construction scenarios. Second, we conducted a well-designed hazard-recognition experiment in which 70 construction workers participated. Their EEG signals were recorded, and after pre-processing, we found that the distributions of the EEG signal power induced by hazardous and safe images conformed empirically with hypothetical brain activities in the constructed conceptual model. This proves that EEG measurements are feasible for observing internal human representations of identification of hazards. Based on this finding, we deduced the perceptual threshold above which the hazard presence was reported, which was then used to predict the binary discrimination choice of individuals from their EEG wavelets for the hazard recognition task.

The EEG-based hazard perceptual threshold was confirmed to be consistent with SDT behavioral indices. Furthermore, we measured the decision confidence of individuals based on their brain activity prior to responding. The degree of confidence was classified into three grades, and the optimal boundary-characterized EEG power for each grade was identified through simulations. Thus, the decision confidence level during hazard recognition can be objectively estimated using an EEG-based BCI instead of traditional self-reporting. Overall, the proposed method offers a promising strategy for construction workers to collaborate with computer vision technologies to identify hazards and demonstrate the potential of BCI as an effective channel for intention prediction and telecommunication. This will lay the foundation for the design of future advanced hazard detection techniques in the collaborative human-machine systems research field.

## Data Availability Statement

All data, models, and codes that support the findings of this study are available from the corresponding author upon reasonable request.

## Acknowledgments

**Funding:** This study was supported by the National Natural Science Foundation of China (No. 51878382).